\documentclass[preprint]{aastex}

\shortauthors{Skillman, C\^ot\'e, \& Miller}
\shorttitle{Abundances in Sculptor Group Dwarfs}

\begin{document}

\title{ISM Abundances in Sculptor Group Dwarf Irregular Galaxies}

\author{Evan D. Skillman\altaffilmark{1}}
\affil{Astronomy Department, University of Minnesota,
    Minneapolis, MN 55455} 
\email{skillman@astro.umn.edu}

\author{St\'ephanie C\^ot\'e}
\affil{Canadian Gemini Office, HIA/NRC of Canada,
5071 West Saanich Rd., Victoria, B.C., Canada, V9E 2E7}
\email{Stephanie.Cote@hia.nrc.ca}

\and

\author{Bryan W. Miller}
\affil{AURA/Gemini Observatory, Casilla 603, La Serena, Chile;}
\email{bmiller@gemini.edu}

\altaffiltext{1}{Visiting Astronomer, Cerro Tololo Inter-American Observatory, 
National Optical Astronomy Observatory, which is operated by the Association 
of Universities for Research in Astronomy, Inc. (AURA) under cooperative
agreement with the National Science Foundation. }

\begin{abstract}

Using the CTIO 4-m telescope, we have obtained optical spectra of 
HII regions in five Sculptor Group dwarf irregular galaxies.
We derive oxygen, nitrogen, and sulfur 
abundances from the HII region spectra.
Oxygen abundances are derived via three different methods (the ``direct'' method,
the empirical method guided by photoionization modeling of McGaugh (1991), and 
the purely empirical method of Pilyugin (2000)) and are compared.  
Significant systematic
differences are found between the three methods, and we suggest
that a recalibration of the empirical abundance scale is required. 
Until differences between these three methods are better understood, the 
issue of the degree of uniformity of the ISM abundances in a dwarf galaxy
cannot be properly addressed.
The N/O ratio for the metal-poor dI ESO~473-G24 of log (N/O) 
$=$ $-$1.43 $\pm$ 0.03 lies
well above the plateau of log (N/O) $=$ $-$1.60 $\pm$ 0.02 found by
Izotov \& Thuan (1999) for a collection of metal-poor blue compact galaxies.
This shows that not all galaxies with 12 $+$ log (O/H) $\le$ 7.6
have identical elemental abundance ratios, and this
implies that the Izotov \& Thuan scenario for low metallicity
galaxies is not universal.
Measurements of the HII regions in NGC~625 yield 
log (N/O) $\approx$ $-$1.25.  Assuming N production by intermediate 
mass stars, this relatively high N/O ratio may be
indicative of a long quiescent period prior to the recent
active burst of star formation. 
The oxygen abundances in the Sculptor Group dIs are in good agreement
with the relationship between metallicity and luminosity observed in
the Local Group dIs.  Taken together the observations show a better
relationship between metallicity and luminosity than between
metallicity and galaxy central surface brightness.
The Sculptor Group dIs, in general, lie closer to 
the simple closed box model evolutionary path than the Local Group
dIs.  The higher gas contents, lower average star formation rates,
and closer resemblance to closed box evolution could all be
indicative of evolution in a relatively low density environment.

\end{abstract}



\section{Introduction}

Low-mass dwarf irregular galaxies provide an important testing ground
for several fundamental questions about star formation, galactic evolution,
and cosmology.  
Due, in large part, to attempts to understand the possible evolutionary
connections between the dwarfs with negligible or extremely low present
star formation rates (the dSph and dE galaxies, hereafter dE galaxies)
and the dwarfs with obvious signs of present star formation (the dIrrs,
blue compact dwarfs, HII galaxies, hereafter dI galaxies), many theorists 
are turning their
attention to the problem of dwarf galaxy evolution (see introduction to
Skillman, C\^ot\'e, \& Miller 2002; hereafter paper 1).  
By comparing the properties of dwarf galaxies in different environments
we may be able to isolate key environmental variables (e.g., local density,
companionship, group vs.\ cluster membership) in order to
constrain these theories.

Dwarf irregular galaxies generally show low metallicities,
(see e.g., Skillman et al.\ 1989a,b), and thus, are ideal laboratories 
for studying the early stages of nucleosynthesis in galaxies and  
constraining the primordial He abundance in the early universe
(see e.g., Lequeux et al.\ 1979; Pagel et al.\ 1992).  
It is generally found that the metallicity of the ISM of dIs
correlates very well with the galaxy's luminosity (Lequeux et al.\ 1979;
Talent 1980; Kinman \& Davidson 1981; 
Skillman, Kennicutt, \& Hodge 1989;
Richer \& McCall 1995; van Zee et al.\ 1997b), but some studies
do not support this (e.g., Hidalgo-Gamez \& Olofsson 1998; 
Hunter \& Hoffman 1999; but see also Pilyugin 2001).
Various explanations of the physical basis for the  metallicity --
luminosity relationship exist, and the simplest are those
based on the idea that dwarf galaxies of decreasing
mass are less likely to retain the recently synthesized
heavy elements returned to the ISM by type II supernovae (e.g., Larson 1974;
Dekel \& Silk 1986; but see also Skillman 1997).

It may be possible to identify the physical process(es) underlying
the metallicity -- luminosity relationship by examining galaxies
with extreme properties (e.g., Kennicutt \& Skillman 2001).  
With that goal in mind, the motivation for 
this study was to measure the ISM abundances in nearby groups of 
galaxies in order to compare with the galaxies from the Local Group.
By comparing galaxies in different environments, it may be possible
to determine environmental effects on the metallicity -- luminosity 
relationship.  For example, Skillman et al.\ (1996) discovered that
stripped spiral galaxies in the core of the Virgo cluster showed
elevated ISM abundances and V\'\i lchez (1995) showed evidence
for elevated abundances in the dwarf galaxies associated with the 
Virgo cluster.

Studies of the relative abundances of nitrogen and oxygen in 
dwarf irregular galaxies potentially provide an independent measurement of
recent average star formation rates.
From a study of the Pegasus dI, Skillman,
Bomans, \& Kobulnicky (1997) found an unusually high N/O ratio
which is interpreted as due to a quiescent star formation history over
the last few 100 million years.  This is in line with the paucity of
HII regions, the red color, and the recent star formation history derived 
from HST stellar photometry (Gallagher et al.\ 1998).
This demonstrates that, in some cases, the N/O abundance can be used as 
a measure of the relative star formation rate over the last 
few 100 million years (Kobulnicky \& Skillman 1998).
Skillman et al.\ (1997) argue that measuring N/O in a sample of dIs
with a large range in colors may allow us to calibrate the time
delay of the N delivery (from intermediate mass stars) relative to
the O delivery (from massive stars).  This is an important parameter
for the interpretation of the abundances of the damped Lyman-$\alpha $
systems (Pettini et al.\ 1995; Lu et al.\ 1998).
Much of the work in this field has concentrated on
the high surface brightness HII galaxies, but there is suspicion
of chemical ``self pollution" of the emission line region by the
current burst of star formation (Kunth \& Sargent 1986;  but see also
discussions in Skillman \& Kennicutt 1993, Kobulnicky \& Skillman 1996, 1997).
This may be less of a concern for the
more modest HII regions which can be observed in nearby dI galaxies;
these systems are not powered by the giant superclusters seen in
more distant galaxies, and so their chemical environments
may be more pristine.  The low present star formation rates observed in
the Sculptor group dwarfs make them ideal objects for testing
the N/O ``clock".

The Sculptor group is the closest group of galaxies beyond our Local Group.
Its properties are reviewed in paper 1 and references in that paper.  
In paper 1 we presented new H$\alpha$ imaging of a sample of Sculptor Group dIs. 
HII regions were detected in eight dwarf irregular galaxies. 
Here we have used the CTIO 4-m to obtain optical spectra of
ten HII regions located in five of these dwarf irregular galaxies.  
These spectra
are used to measure chemical abundances, which are compared to the 
HII region chemical abundances available for Local Group dIs.

\section{Spectrophotometric Observations}

  Spectra were taken with the cassegrain R-C spectrograph ($f$/7.8) on
the CTIO 4-m telescope on the evening of 6 September 1997 and
the first half of 7 September 1997.  High humidity and precipitation
prevented observations on the second half of 7 September 1997 and
8 September 1997.
We used the Blue Air Schmidt camera with a
Loral thinned 3K $\times$ 1K format CCD with 15 $\mu$ pixels
as the detector. A  527 line mm$^{-1}$ grating (KPGL3)
resulted in a dispersion scale of 1.91 \AA\ 
pixel$^{-1}$ and a spatial scale of 0.50\arcsec pixel$^{-1}$.  
A uv (WG360; 3600 \AA ) blocking filter was used to 
suppress second order contamination in the red.
Useful data were collected over the wavelength range of
3600 -- 7100 \AA .
Observations were obtained with a 1.3\arcsec\ wide slit
observing at low air mass and near the parallactic angle in order to avoid
problems of differential atmospheric refraction (cf.\ Filippenko 1982).
The projected slit length is slightly larger than 3\arcmin .

  Bias frames, dome flats, 
twilight sky flats, and He-Ne-Ar comparison exposures were taken at the 
beginning and end of the nights.  On the first night, three standard stars 
from the list of Oke (1990) were observed (2 $\times$ LTT 9491, 
2 $\times$ Feige 110, G158-100).  
LTT 9491 and Feige 110 were observed on the second night. The standard stars 
were observed with a slit width of 6$^{\prime\prime}$ in order to avoid
any effects of differential atmospheric refraction.

 Observations of a total of ten HII regions in five different galaxies
were obtained as follows: 
 2 $\times$ 1800s observations were made of E347-G17 \#5 \& \#10 (see paper 1) 
at an average airmass of 1.2 with an E-W slit position angle,
 3 $\times$ 1800s observations were made of E348-G09 \#3 (paper 1)
at an average airmass of 1.1 with an E-W slit position angle,
 4 $\times$ 1200s observations were made of E471-G06 \#2 (as numbered
by Miller 1996) at an average airmass of 1.4 with an E-W slit position angle,
 4 $\times$ 1800s observations were made of E473-G24 \#2 \& \#4 (paper 1) 
at an average airmass of 1.02 with an E-W slit position angle, and
 3 $\times$ 1800s observations were made of NGC~625 \#5, \#9, \#18, and \#21 
(paper 1) at an average airmass of 1.06 with a position angle
of $-$98\degr .
 
Standard reduction procedures
were followed using the programs available within the IRAF\footnotemark\ system.
\footnotetext{IRAF is distributed by the National Optical Astronomy
Observatories, which are operated by the Association of Universities for
Research in Astronomy, Inc., under cooperative agreement with the National
Science Foundation.}
The standard star observations from the two nights were reduced independently 
and compared.  Because no systematic differences could be detected, both
night's data were combined and analyzed together (i.e., using the same 
sensitivity function and the same extinction law).  Because G158-100
is a relatively red star (spectral type sdG) the variation between
its sensitivity calibration and the other two bluer stars indicated
at what wavelength the spectra of the blue stars were beginning to 
be affected by second order contamination ($\sim$ 6860 \AA , although
the atmospheric ``B'' band due to O$_2$ absorption confuses the 
precise onset of the effect).  Note that second
order contamination is found well before the $\lambda$7200 that one
derives by simply doubling the $\lambda$3600 wavelength of the 
second-order blocking filter.  This is simply because the $\lambda$3600
refers to the 50\% point of the filter, and the profile of the 
transmission of the filter is far from step-like.  Beyond this
wavelength only G158-100 was used for determining the 
sensitivity curve (the flux of G158-100 is roughly 2 magnitudes weaker
at $\lambda$3500 compared to $\lambda$7000, so this should provide a
good calibration out to the red limit of our observations at $\sim$ 
$\lambda$7100).  
This means that for emission lines between $\sim$ 6860 \AA\ and $\sim$ 
$\lambda$7100 \AA , the fluxes of the emission lines should not be 
affected, but that  the continuum of the HII region targets may be 
contaminated by second order light (thus reducing the emission line
equivalent widths).  In principle, this is only important for using the 
[He~I] $\lambda$7065 line to derive a helium abundance if corrections for 
underlying absorption are applied on an equivalent width basis. 
Because the standard stars were observed over a large range in airmass, 
during the flux calibration stage, an extinction law could be derived from
the standard star data.  This agreed to within a few percent with the 
extinction law derived for the CTIO observatory and supplied within IRAF.  

  Extracted spectra are shown for the brightest HII region in each of the
five galaxies in Figures 1a-e.
Emission lines were identified, and fluxes and errors were measured
following the method of Skillman \& Kennicutt (1993; SK93).  
Values of the logarithmic extinction at
H$\beta$, C(H$\beta$), were derived from the error weighted
average of determinations from the H$\alpha$/H$\beta$, H$\gamma$/H$\beta$,
and H$\delta$/H$\beta$ ratios while simultaneously solving for the effects
of underlying stellar absorption, EW(HI-abs), (assumed to be equal in equivalent 
width for all four Balmer lines).  
We assumed the intrinsic case B Balmer line ratios calculated by Hummer \& Storey 
(1987), and   used the reddening law of Seaton (1979) as parameterized by Howarth 
(1983), assuming a value of R $=$ A$_{V}$/E$_{B-V}$ $=$ 3.2.  
The assumed reddening law has no associated uncertainty.
Uncertainties in C(H$\beta$) and EW(HI-abs) were determined from Monte 
Carlo simulations (Olive \& Skillman 2001).  Figure 2 shows an example solution
for one of the observed HII regions.  Note that the errors derived in this
way are often significantly larger than those derived in the literature
by either assuming a value for the underlying absorption or derived 
from a $\chi ^2$ analysis without a Monte Carlo analysis of the errors.
When the Balmer lines were corrected for underlying absorption, 
the higher numbered Balmer lines (H9, H10, H11, and H12) were not assumed to 
have identical values in terms of equivalent width, but were assumed to
have values of 0.80, 0.65, 0.50, and 0.40 respectively of the lower numbered
Balmer lines, guided by the low metallicity, instantaneous burst models of 
Gonz\'alez-Delgado, Leitherer, \& Heckman (1999).
Note that all of the Sculptor Group dIs lie at Galactic latitudes more
negative than $-69$, and thus, Galactic extinction 
along the lines of sight to the H~II regions should be negligible
(e.g., the Galactic foreground E(B-V) for the Sculptor dwarf irregular galaxy, 
which lies near the middle of the projected distribution, is calculated to
be 0.012 mag by Schlegel et al.\ 1998). 
The corrected fluxes for the ten HII region spectra,
relative to H$\beta$, are listed in Table 1. 

\section {The Chemical Abundances in Five Sculptor Group
Dwarf Irregular Galaxies}

\subsection{The Oxygen Abundance}

Oxygen abundances are derived for the HII regions by three different methods:
the ``direct'' method (cf.\ Dinerstein 1990, Skillman 1998), the empirical
method of McGaugh (1991), which is guided by the use of photoionization models,
and the purely empirical method of Pilyugin (2000).
A reliable measurement of the electron temperature of the ionized gas
is necessary for the ``direct'' conversion of emission line strengths 
into ionic abundances.  
In most cases of metal poor HII regions, this
is done by measuring the [O~III] $\lambda$4363/($\lambda$4959 + $\lambda$5007)
ratio.  We have measured the temperature sensitive [O~III] $\lambda$4363
line in HII regions in four of the five galaxies (6 of the 10 observed
HII regions).
For deriving the temperature in the [O~III] zone, and all subsequent
abundance calculations, we use the emissivities determined from
the five-level atom program of Shaw \& Dufour (1995).
The error in T$_{\rm e}$(O$^{++}$) is derived from the emission line uncertainties
and does not include terms for uncertainties in the atomic data nor for
the presence of temperature variations within the [O~III] zone.  
Following SK93, we use the formula derived by 
Pagel et al.\ (1992) for estimating the temperature in the low ionization zone
based on the models of Stasi\'nska (1990):
\begin{equation}
{\rm T_e(O^+) = 2 (T_e^{-1}(O^{++}) + 0.8)^{-1}}
\end{equation}
where T$_{\rm e}$ is the electron temperature in units of 10$^4$ K.
We also assume T$_{\rm e}$(N$^+$) and T$_{\rm e}$(S$^+$) are equal to
T$_{\rm e}$(O$^+$) (cf.\ Garnett 1992).  In some cases, the error in 
T$_{\rm e}$(O$^+$) is unrealistically small if calculated directly 
from the error in T$_{\rm e}$(O$^{++}$),
and a lower limit of 500 K in the error of T$_{\rm e}$(O$^+$) is assumed. 
From the [S~II] lines we can calculate electron densities 
which are generally in the low density regime, (less than 150 cm$^{-3}$,
1 $\sigma$). 
Ionic abundances for O$^{++}$ and O$^+$ were then computed
from the emission line ratios and we  
derived the oxygen abundance $\equiv$ O/H = O$^{++}$/H$^+$ + O$^+$/H$^+$
for all regions with $\lambda$4363 observed.
A summary of ionic and total abundances via the direct method is shown in Table 2.

In the absence of a direct measurement of the electron temperature,
photoionization models can act as a guide to determining oxygen abundances
(cf.\ Edmunds \& Pagel 1984).
This is discussed specifically in the case of low abundance H~II
regions by Skillman (1989) where it was demonstrated that, due to a low 
sensitivity to the hardness of the radiation field, the [O~II] and [O~III] 
line strengths can be combined to uniquely determine the ionization
parameter and an ``empirical'' oxygen abundance.  McGaugh (1991)
has produced a grid of photoionization models and recommended
using the ratios of ([O~III] + [O~II])/H$\beta$ $\equiv$ 
($\lambda$4959 + $\lambda$5007 + $\lambda$3727)/H$\beta$ $\equiv$ R$_{23}$ and
[O~III]/[O~II] $\equiv$ ($\lambda$4959 + $\lambda$5007)/$\lambda$3727
(hereafter O32)  to
estimate the oxygen abundance.  Figure 3 shows this diagnostic diagram
with the models of McGaugh (1991) plotted and the observations of
all ten observed Sculptor Group dI HII regions plotted.
We have used the McGaugh models to determine empirical oxygen abundances
and associated statistical errors, and have listed them in Table 3.  

Pilyugin (2000) has proposed a new calibration of the empirical
oxygen abundance scale.  After rearranging terms, Pilyugin's calibration is:
\begin{equation}
12 + log(O/H) = 6.35 
+ 3.19\ log\biggl(\frac{I(\lambda 4959 + \lambda 5007 +\lambda 3727)}{I(H\beta )}\biggr)
- 1.74\ log\biggl(\frac{I(\lambda 4959 + \lambda 5007)}{I(H\beta )}\biggr)
\end{equation}
We have added the oxygen abundances calculated via Pilyugin's calibration
to Table 3, and compared Pilyugin's calibration to that of McGaugh in
Figure 4.  There are two significant and systematic differences between
these two empirical calibrations for low metallicity HII regions.
First, note that Pilyugin's calibration extends to larger values of
log (R$_{23}$).  This solves the well known problem that McGaugh's grid 
does not cover the highest values of log (R$_{23}$) that are observed.
Because Pilyugin's calibration is entirely empirical, it covers the
full range of the observations.  The second significant difference
is the change in slopes of the calibrations at low values of log (O32).
Since most of the calibrating observations used by Pilyugin had high
values of log (O32), the extrapolation of his calibration into the
lower log (O32) regime is subject to very large uncertainty.  Note, 
for example, the 0.34 dex difference between M91 and P00 for the low
excitation HII regions NGC~625 \#21.
Because McGaugh's calibration is guided by photoionization models in 
this regime, it is probably the more accurate there. 

Figure 5 shows the differences between the three oxygen determinations
given in Table 3 as a function of both log (O32) and log (R$_{23}$).
When the differences are plotted as a function of log (O32), the 
systematic difference between the M91 and P00 calibrations is obvious.
As could also be seen in Figure 4, in some cases, the 
empirical oxygen abundances are in good agreement with the direct
oxygen abundances, with a slight bias towards lower oxygen
abundances from the direct measurements (i.e., the differences for  
ESO~471-G06 \#2, ESO~473-G24 \#2, and NGC~625 \#18 all fall between 
-0.03 and -0.15 dex).  The small bias could be indicative of any of
a number of different effects.  One possibility is the
breakdown of the assumption of a uniform electron temperature (see 
Peimbert, Peimbert, \& Luridiana 2002 and references therein).
A second related possibility is the presence of a second heating 
source in addition to photoionization (Stasi\'nska, Schaerer, \& 
Leitherer 2001).  A third possibility is that radiation fields are
much softer that those considered by the models (McGaugh 1991
shows that this is a relatively small effect - compare the two models
plotted in Figure 3 - but Stasi\'nska et al.\ 2001 caution that 
use of only zero-age stellar models does not probe a large enough
range in input stellar models).

However, there are some truly discrepant results which do require attention.  
ESO~347-G17
\#5 shows a difference of $-$0.35 dex for the M91 calibration and $-$0.32
for the P00 calibration.  On the other hand, the empirical abundances
for HII region ESO~347-G17 \#10 differ by 0.31 dex. 
The brightest two HII regions for NGC~625 also show significant
discrepancies.  The empirical abundances
for NGC~625 \#5 and \#9 are between $-$0.19 to $-$0.35 dex below those 
of the direct oxygen abundances.  It is interesting that the 
empirical oxygen abundances for the three brightest NGC~625 HII regions
all agree within a range of 0.24 dex, 
while the direct oxygen abundances cover a range of 0.23 dex, but
the order in oxygen abundance is different for each method.  

These significant differences lead us to two important conclusions:
First, we feel that both of the existing empirical oxygen abundance
calibrations for metal poor HII regions have deficiencies and that
the problem of calibrating the empirical oxygen abundances should
be revisited with a much larger database consisting of observations
spanning a large range in both log (R$_{23}$) and log (O32).
Second, the uncertainties quoted when using empirical oxygen abundances
probably have been underestimated in the past. McGaugh (1991) estimated
an uncertainty of 0.05 dex for low metallicity HII regions.  However,
this was based on a comparison with the abundances of Campbell (1988).
Although the observations used by Campbell did include measurements of
the [O~III] $\lambda$4363 line, Campbell's abundances were based on 
photoionization model fits to only the bright [O~II] and [O~III] lines.
Thus, this estimate of the uncertainty is specious, and only shows 
the range in results of different photoionization model fits to the
bright lines.  Pilyugin (2000) claims that his calibration is as 
accurate as using the direct method, and this could be true for
the regime of high values of log (O32), but this is almost
certainly not true for lower values of log (O32).

All of the above are important to the question of whether abundances
in dI galaxies are uniform or if there are real variations.   
Generally, dwarf irregular galaxies are observed to 
have uniform abundances (Kobulnicky \& Skillman 1996, 1997;
and references therein).  However, given that usually only a handful
of HII regions are observable in a dI and that only a percentage of
them have [O~III] $\lambda$4363 bright enough to be well measured,
it is then problematic to seriously test the level of abundance
variations.  For example, are there abundance variations in NGC~625?
Both the direct methods and the empirical methods would indicate 
that there are, but none of the methods are in agreement.  Because the
direct method is subject to several systematic uncertainties ($\lambda$4363
is an inherently weak line and, at low velocities, is liable to possible 
contamination by terrestrial Hg emission; the assumption of a single
temperature in the O$^{++}$ zone is probably not correct; the assumption of 
a single relationship between the temperature in the O$^{++}$ and O$^{+}$
zones is certainly only a first order assumption), a measurement of 
$\lambda$4363 should not be taken as prima facie evidence of an accurate 
oxygen abundance. 

Nonetheless, given the present observations, is it possible that there
are measurable differences in the oxygen abundances in NGC~625?
Looking at the direct measurements, the two brightest HII regions
(\#5 and \#9) are in excellent agreement, but the third HII region (\#18)
shows an electron temperature which is about 2,000 K higher, and thus
an oxygen abundance about 0.2 dex lower.  Is this result to be taken
at face value?  While the $\lambda$4363 line is relatively weak,
there is no contaminating Hg emission, and the associated errors result
in only a 0.04 dex error in the oxygen abundance.   Is it possible that
the brightest two HII regions have experienced some self-pollution?
There is evidence from the spectra that the ionizing clusters in 
the two brightest HII regions are old enough that the most massive
stars have evolved off the main sequence phase.
Figure 6, a close-up in the region of $\sim$$\lambda$4700 in region \#5,
shows the presence of broad He II emission usually associated with the
winds of massive early-type stars with ages of at least 2 million years.  
The equivalent width of the $\lambda$4686
feature is 5.5 $\pm$ 1.5 \AA, and there appears to be no evidence in 
the spectrum for the Wolf-Rayet emission features at $\lambda$4640 (N~III), 
$\lambda$4650 (C~III/C~IV), or $\lambda$5808 (C~IV) at the 0.1 \AA\ level.
Broad He II $\lambda$4686 emission is also detected in region \#9, although
at a much lower level (0.7 \AA\ EW).  

The lack of the W-R features associated with C and N is indicative 
of low metallicity (Maeder, Lequeux, \& Azzopardi 1980; Arnault, 
Kunth, \& Schild 1989), and, in the detailed models, these 
features tend to dissappear at metallicities of around one-tenth of
the solar value (Kr\"uger, et al.\ 1992; Schaerer \& Vacca 1998).
Although it is undeniably desperate to attempt to infer abundances
from the presence or absence of W-R features, the absence of these W-R
features may favor the lower abundances (for the stars), implying that 
the nebular abundances could have been enhanced by the return of newly 
produced oxygen.  However, the similar N/O abundances seen in the 
brightest 3 HII regions would argue against the recent enhancement
(see next section).   

Additionally, the empirical abundances give the opposite result in that 
the two brightest HII regions are lower in oxygen abundance than the
two fainter HII regions.  Basically, in this case, it appears to be
difficult to pin down the abundances at better than the 0.2 dex level,
which is approximately the range of the measurements.  Given the interest
in the problem of ``self-pollution'' and its role in the chemical 
evolution of dwarf galaxies (e.g., Kunth \& Sargent 1986, Tenorio-Tagle 
1996, Recchi, Matteucci, \& D'Ercole, 2001) a clearer view of the
observational situation is probably warranted. 

\subsection{The N/O Abundance Ratio}

\subsubsection{Theories of the N/O Ratios in Dwarf Galaxies}

A rather substantial review of the observational and theoretical aspects 
of the N/O abundance ratio is given in Henry, Edmunds, \& K\"oppen (2000).
Here we list just a few important highlights.
While the production of O in galaxies is dominated by nucleosynthesis
in massive stars and relatively prompt return to the ISM via supernovae
type II explosions, the dominant processes leading to the enrichment of N in
galaxies is still debated.  In the more massive
spiral galaxies, at higher metallicities, N/O increases roughly linearly
with O/H, indicative of a secondary production mechanism
dominating N production (Vila-Costas \& Edmunds 1993).
This secondary production of N is expected to come from C and O in the 
CNO cycle in intermediate mass stars, with subsequent release into the 
ISM via red giant winds and planetary nebula.
In contrast, dwarf irregular galaxies also show relatively constant N/O 
at low metallicities with increasing upward scatter at higher metallicities
(Garnett 1990, Thuan et al.\ 1995, Kobulnicky \& Skillman 1996).  A 
constant value of N/O over a large range in metallicity could be taken as
evidence of primary production of N (Pagel 1985).

  Three processes leading to primary N production have been suggested.
Renzini \& Voli (1981) showed that increasing the convective scale length
over the pressure scale length resulted in freshly synthesized $^{12}$C
being brought up to the convective envelope whereupon it can by converted
to N via the CNO cycle.  This ``hot bottom burning" mechanism
(Iben \& Renzini 1983) will be most effective in the mass range of 4
to 5 solar masses.
Timmes, Woosley, \& Weaver (1995) and Woosley \& Weaver
(1995) have suggested that primary N might be produced in massive stars
(heavier than 30 solar masses) of low metallicity.  Again, if convection is
enlarged beyond the standard models, a convective, helium burning
shell penetrates into the hydrogen burning shell, and freshly
synthesized C is converted into N.
Recently, Meynet \& Maeder (2002) have shown that when the effects of
stellar rotation are included, stars of metallicities lower than about
1/5th of solar are capable of producing a large amount of primary N 
due to enhanced shear mixing.  Their mechanism is important for stars
with masses between 2 and 7 M$_\odot$. 
Note that the first and third primary N production processes would have 
comparable delivery times of a few 100 million years, while the second 
process would be of order 10 million years.

  Garnett (1990) discussed the possibility that the spread in the
values of N/O at a given O/H as measured in dwarf irregulars could
be attributed to a delay between the delivery of O and N to the
ISM (as proposed by Edmunds \& Pagel 1978).  If a dwarf galaxy
experiences a dominant global burst of star formation, then the ISM O
abundance will increase after roughly 10 million years, with a
resulting decrease in N/O.  (This assumes that the O is immediately incorporated
into a visible component of the ISM, i.e., the warm phase, which is
argued against by Kobulnicky \& Skillman 1997.) 
Then, over a period of several
hundred million years, the N/O abundance ratio will increase at
constant O/H (given the absence of a subsequent burst of star
formation in that time interval).
Under these assumptions
(dominant bursts of star formation separated by quiescent periods
and delayed N delivery to the ISM) the N/O ratio becomes a clock,
measuring the time since the last major burst of star formation.
Low values of N/O imply a recent burst of star formation, while
high values of N/O imply a long quiescent period.  If it takes
the O a few tens of millions of years to be incorporated
into a visible component of the ISM, then the ``clock'' will still be
effective, but if it takes a few 100 million years for this to 
take place, then bursts are unlikely to drive the N/O variations.
Since the N/O variations are observed, and appear to be real, this
suggests that the O mixing time scale is short compared to the N
release timescale. 

Henry et al.\ (2000) have presented models which produce a plateau in 
N/O at low values of O/H with a steep rise in values of N/O at higher 
values of O/H by assuming both primary and secondary production of N 
by intermediate mass stars.  
For the case of bursting galaxies, they note that as long as the period
between the bursts is longer than about 250 million years, then the
N from the intermediate mass stars will have sufficient time to 
be released and mixed, so that even these galaxies should lie close
to the trend established by their slowly decreasing star formation 
rate models.  In order to explain the scatter in N/O at higher values 
of O/H, they favor a scenario of fast enrichment of N by Wolf-Rayet 
stars or luminous blue variable stars.

Larsen, Sommer-Larsen, \& Pagel (2001) have recently investigated the 
problem, and, based on simultaneous constraints from N, O, and He
abundances and the relationship between metallicity and gas mass
fraction, they conclude that the scatter in N/O at given O is
best understood by a double-bursting mode of star formation.
In this scenario, a burst of star formation triggers a second burst
with a delay of roughly 30 million years. 

Hopefully, observations of abundance ratios in several HII regions
in dwarf galaxies that are close enough to construct recent star
formation histories (like the present sample of Sculptor dIs)
will allow us to distinguish between these 
different interpretations.
 
\subsubsection{Our Results and Implications}

For the present discussion, we will concentrate on the higher quality 
spectra (those with $\lambda$4363 detected).
In order to convert the N$^+$/O$^+$ ionic abundance ratio into a total
abundance ratio (N/O), we need to account for any differences in
ionic fractions.  We adopt the standard assumption:
N$^{+}$/O$^{+}$ $\equiv$ N/O (Garnett 1990).
The N/O ratios are given in Table 2.

Figure 7 presents a comparison of the N/O and O/H in the Sculptor Group
dIs with the collection of dwarf irregular galaxies and H~II galaxies
assembled by
Kobulnicky \& Skillman (1996; see their Table 5 and Figure 15 for
identification of individual points).  Also added to the figure are
the low surface brightness dwarfs of van Zee et al.\ (1997b) and
the low metallicity BCDs of Izotov \& Thuan (1999, hereafter IT99).
For three of the four galaxies, the results, considering errors,
are relatively close to the average value of N/O $=$ $-$1.47 determined
for dwarf irregular galaxies by Garnett (1990).  
More recent studies have found similar averages.
Thuan et al.\ (1995) found an average of log (N/O) = $-$1.53
$\pm$ 0.08 for a sample of 14 blue compact galaxies (BCGs). 
van Zee et al.\ (1997b) found an average of log (N/O) = $-$1.56 $\pm$ 0.11
for their sample of low surface brightness dwarfs.
In the present sample, only NGC~625 deviates
significantly, with higher N/O ratios in all three of the observed
HII regions.   Nonetheless, each of the Sculptor Group galaxies are
interesting in their right, and we will discuss them individually. 

ESO~473-G24 presents a particularly interesting case.
IT99 have derived an average of log (N/O) $=$ $-$1.60 $\pm$
0.02 for a sample of 6 BCGs with 12 $+$ log (O/H) $\le$ 7.6.
Based on this, IT99 propose a scenario in
which all galaxies with 12 $+$ log (O/H) $\le$ 7.6 are undergoing 
their first ever burst of star formation and their elemental abundance
ratios represent pure primary production.  
ESO~473-G24 does not follow this pattern.  For this
galaxy, the empirical oxygen abundances for both HII regions and the
direct oxygen abundance for the brighter HII region are all at or below
12 $+$ log (O/H) $=$ 7.6.  However, log (N/O) is measured to be
$-$1.43 $\pm$ 0.03.  
The present observations
of ESO~473-G24 show that not all galaxies with 12 $+$ log (O/H) $\le$ 7.6
have identical N/O abundance ratios.  

In this regard, ESO~473-G24 is similar to DDO 154.
Kennicutt \& Skillman (2001) find that DDO 154 (12 $+$ log (O/H) $=$
7.67 $\pm$ 0.05) is elevated in N/O ($=$ $-$1.44 $\pm$ 0.05) relative 
to the extremely narrow 
plateau defined by the low metallicity BCDs of IT99.  However, the 3.2 
$\sigma$ difference is not striking and would be even less so if not for the
very small dispersion (0.02 dex) in N/O for the IT99 sample.
Kennicutt \& Skillman (2001) have pointed out that the observed narrow
dispersion in the IT99 observations itself is very difficult to understand 
given the associated uncertainties in reddening corrections, the 
ionization correction factor, and the estimated temperature in the O$^+$ 
zone.

It is already well known
that the Local Group dwarf galaxy Leo A has an oxygen abundance of
12 $+$ log (O/H) $\approx$ 7.3 (Skillman et al.\ 1989; 
van Zee, Skillman, \& Haynes 2002), 
and yet shows the bulk of its
stars were formed over the last few Gyr (Tolstoy et al.\ 1998).
Recently, Dolphin et al.\ (2002) have shown evidence for an ancient
(age $\sim$ 10 Gyr) stellar population in Leo A.  
This implies that the IT99 scenario is certainly not universal, and
casts doubt on the conclusion that the N observed in low metallicity 
galaxies was all produced recently in massive stars.

The high N/O ratio seen in all three of the HII regions of NGC~625
is notable in that it lies close to the upper envelope of values
observed in dIs and blue compact galaxies.  Although the N/O ratio is not
as high as the extreme of log (N/O) $=$ $-$0.85 seen in one region of
NGC~5253 (Kobulnicky et al.\ 1997 and references therein), it does
lie in a part of the diagram populated only by blue compact
galaxies.  There are alternative possible explanations for the high
N/O in NGC~625.  On one hand, it may be that a long quiescent period is
required before a galaxy can experience such a strong burst of
star formation like that seen in NGC~625.  In this case, the N/O
ratio would reach its maximum as the delayed nitrogen production
has the opportunity to come to completion.  The alternative is
that something similar to the case of NGC~5253 has occurred, where
prompt enrichment of N is observed (i.e., N released from the winds
of the massive stars exciting the HII region is responsible for
``polluting'' the local gas to a higher value than the surrounding
ISM).  The similarity of the N/O
ratios in the different HII regions of NGC~625 imply that the former
possibility is more likely (in NGC~5253 the surrounding ``normal''
HII regions have log (N/O) $=$ $-$1.3).  An HST study of NGC~625
is now underway (Cannon et al.\ 2002), and it will be interesting to
compare the estimates of the ages of the exciting cluster stars 
in each of the observed HII regions.   

If the hypothesis of delayed N production holds up as the explanation
for NGC~625, it would join other observational studies which point
to N production predominantly in intermediate mass stars.
Based on the observation of high N/O in the Pegasus dwarf irregular galaxy,
Skillman et al.\ (1997) have proposed that dwarf galaxies with low
current star formation rates may be a good location to look for
enhanced N/O.
From HST observations, Gallagher et al.\ (1998) found that the star
formation rate in the Pegasus dwarf irregular galaxy has been relatively
low for the last few 100 million years, supporting the 
hypothesis of high N/O due to delayed N production.  
Although ESO~473-G24 and ESO~471-G06 lack 
detailed studies  of their stellar populations, they both have extremely low
current star formation rates for their total gas content resulting
in a gas depletion time scales of 65 Gyr and 110 Gyr, respectively
(cf.\ paper 1, Table 3).
Their N/O ratios appear to be at least consistent with 
the proposal that we are seeing the effects of completed delayed N 
production in these relatively low metallicity systems.

In this regard, the galaxies with relatively {\it low} values of N/O may
hold the key to understanding the scatter in N/O.  If a galaxy is
observed a sufficient amount of time after a burst to allow the O
to be released and mixed, but not too long such that N production
hasn't reached its peak, a low N/O is predicted by the delayed
delivery hypothesis.  
Skillman, Terlevich, \& Melnick (1989) suggested that this might be
the case in NGC~6822, and Gallart et al.\ (1996) showed evidence for
a significant increase in the star formation rate during the period
100 to 200 million years ago.  With its low N/O ratio, ESO~347-G17 
may present another target worthy of constructing a recent star 
formation history.  Although it appears to lie at the back of the
Sculptor Group, with a nominal distance of 7 Mpc (paper 1), it is
close enough that a reliable recent star formation history can
be constructed from Hubble Space Telescope observations
(e.g., Dohm-Palmer et al.\ 1997, 1998). 

If the evidence presented here favors the delayed N production as
an explanation of the observed scatter in N/O, then does this 
compare with regard to Henry et al.'s observation that the
data are better understood in terms of N contamination and not O 
contamination?
They favor N contamination because the majority of excursions away
from their trend line are upward (increasing N), not downward
(increasing O).
We have two observations to add here. First, in the regime of
7.5 $\le$ 12 + log (O/H) $\le$ 8.3, the situation is not so clear,
with large numbers of excursions both above and below their
model trend line.
Second, the observations in this regime are heavily weighted toward
blue compact galaxies.  If typical burst durations are closer to 
100 million years than to 10 million years (e.g., NGC~1569,
Vallenari \& Bomans, 1996; Greggio et al.\ 1998) then the 
observations may present a biased sample and the mean N/O in
this range may be higher than the mean in the sample.  
Henry et al.\ (2000) preface their conclusion with the caveat that their
interpretation assumes no selection effects in their compilation
of observations.

\subsection{Other Relative Abundances}

The log (S/O) abundance ratios in Table 2 vary between $-$1.36 and $-$1.55.
This is comparable to the range of $-$1.45 to $-$1.60 observed in
low metallicity BCDs by Thuan et al.\ (1995).  These all support the
observation of Garnett (1989) that the S/O ratio shows no dependence
on metallicity and likely reflects the relative yields of S and O
in massive stars.

Since the value of the 
helium abundance requires an accuracy of a few percent to be of
use in constraining big bang nucleosynthesis, and since different
methods of analysis produce variations of order a few percent
(Peimbert, Peimbert, \& Ruiz 2000; Olive \& Skillman 2001), we
have not calculated He abundances for these observations.  Most
of the HII regions are low surface brightness, and therefore will
not provide a useful constraint on the primordial He abundance.
The brightest region in NGC~625 is the exception, with very high s/n 
in even its faint HeI lines (e.g., $\lambda$4026, 7065).  
We defer a detailed analysis of the helium abundance in NGC~625 to
a later paper.  Note that with an electron temperature of $\sim$ 11,000K,
the HII region in NGC~625 lies in the regime where temperature
fluctuations will be very small (Peimbert, Peimbert, \& Luridiana 2002).
Thus, this object may be very useful in this regard.

\section{The Metallicity -- Luminosity Relationship for Dwarf Irregular
Galaxies}

  Since the nebular abundance studies of the Magellanic Clouds in the
mid 1970's (e.g., Peimbert \& Torres-Peimbert 1974, 1976),
it has been suggested that there might be a correlation between
galaxian mass and the metallicity of the interstellar medium.
This has been supported by observations of 
H~II regions in irregular galaxies by
Lequeux et al.\ (1979), Talent (1980), Kinman \& Davidson (1981), 
Skillman et al.\ (1989), Richer \& McCall (1995), Miller (1996),  
and van Zee et al.\ (1997b). 
There have been suggestions that the fundamental relationship may not be 
between mass and metallicity, but between surface density and metallicity  
(Mould, Kristian, \& DaCosta 1983; Bothun \& Mould 1988;
Edmunds \& Phillips 1989).
Based on observations of two dEs in the M81 group, 
Caldwell et al.\ (1998) concluded that luminosity, and not surface 
brightness, is the key parameter determining metallicity for the dEs.

   Kennicutt \& Skillman (2001), drawing on observations from the literature,
concluded that luminosity, and not surface brightness, is probably the 
important parameter for determining metallicity in dIs also.  In Figure 8,
we compare our new observations of the Sculptor Group dIs to the 
compilation of Local Group dIs by Mateo (1998).  We have added in the
oxygen abundance for the Sculptor Group dI A143 $=$ ESO~245-G05
(see paper 1) reported by Miller (1996) as 12 + log (O/H) $=$ 7.97.
Once again, while the known correlation in O/H versus L is clearly shown, 
the correlation is much weaker (or absent) for O/H versus surface brightness. 
It would appear that this comparative study supports luminosity 
as the important parameter.
  
  It would be best to place this conclusion on a firmer statistical base,
but the sample size is small and non-uniform.  In order to invetigate only
the scatter in Figure 8, we have assumed errors of 0.1 dex in all of the
abundance measurements and 0.2 mag in the luminosities and surface brightness.
Then performing a linear least squares fit to both samples yields nearly
identical values of $\chi ^2$ (60), even though the sample with luminosities
is larger (19 versus 13).  Running the same test on only the galaxies
with both luminosities and surface brightnesses yields a factor of two
lower $\chi ^2$ for the regression against luminosity (confirming
what the eye sees in Figure 8).  We repeat that it would be best to 
perform this type of test with a larger database with uniform error
estimates.

\section{The Chemical Evolution of Dwarf Irregular Galaxies}

There has been a long standing debate concerning the role of galactic winds
in the evolution of dwarf galaxies (see Skillman 1997 for a review).  
One aspect of this debate concerns
the yields derived from the observations of dwarf galaxies.  
Often, the calculated yields are significantly lower than theoretically 
expected yields, and this has been taken as good evidence of losses 
due to metal enriched winds (Matteucci \& Chiosi 1983).

In the simple closed box model with instantaneous recycling (Searle \& 
Sargent 1972), the gas phase abundance ($Z$) is related directly to the
baryonic gas mass fraction ($\mu$) as:

\begin{equation}
Z\ =\ y\ ln ({1 \over \mu})\ =\ y\ ln({M_{\rm stars} \over M_{\rm gas}} + 1 ),
\end{equation}

\noindent where $y$ is the elemental yield.  In principle, the heavy
element yield can be derived from the present observations of 
the dI galaxies.  This requires various assumptions, some of which
are particularly tenuous  (see, e.g., Kennicutt \& Skillman 2001).
For example, in order to derive the total gas mass, the HI mass is 
corrected for the presence of He (which is
quite reasonable), and the fraction of gas mass in the molecular and hot
phases is assumed to be small (probably less secure).
It is also generally assumed that the entire HI disk has the metallicity 
as measured in the HII regions (which is quite uncertain, especially
for very extended HI disks).
And, finally, the conversion from optical luminosity to stellar mass
carries a fair degree of uncertainty.  
If the assumptions of the simple closed box model are not appropriate,
then the yield derived in this way is referred to as the ``effective'' yield.

In the study of DDO 154, Kennicutt \& Skillman produced a simple diagnostic
diagram of M(HI)/L(B) versus log (O/H) which, roughly, corresponds to 
gas mass fraction versus metallicity.  There, it was shown that DDO 154
was consistent with a simple closed box model of chemical evolution with
an effective yield ($ y = 4.2 \times 10^{-4}$) that was close to that 
predicted theoretically.
van Zee et al.\ (1997a) showed this to be true for a number of LSB dIs 
and questioned the notion that these galaxies lose a large fraction of
their gas mass due to galactic winds.
In Figure 9, we present a similar graph comparing the Sculptor Group dIs with 
the Local Group dIs from the compilation of Mateo (1998)\footnotemark, 
and the LSB dwarf irregulars from van Zee et al.\ (1997a).  
Superimposed on this
comparison is the chemical evolution track for a simple closed
box model derived for the very gas rich, low surface brightness
dI DDO 154 (Kennicutt \& Skillman 2001).  For this model
an effective oxygen yield of 4.2 $\times$ 10$^{-4}$, or roughly 50\%
of the solar value was used\footnotemark.
This is close to favored values for the
theoretical or ``true'' yield (c.f.\ Maeder 1992), and was derived
from estimates of the total stellar mass, the total gas mass, and the
assumption of constant metallicity throughout the gas disk for 
DDO 154.  An additional model with a factor of 3 lower yield 
(1.4 $\times$ 10$^{-4}$) has also been plotted in Figure 9.
Most of the Sculptor Group dIs and LSB dIs lie between these two models.
\footnotetext{The value of M(HI)/L(B) for SagDIG in Table 4 of Mateo (1998) of
8.6 is in error, and we use the corrected value of 0.86 here.}
\footnotetext{For the recently favored value of the solar abundance of
4.9 $\times$ 10$^{-4}$ (Allende Prieto, Lambert, \& Asplund 2001), this
is equivalent to 86\% of the solar value.} 

There are two notable features to the distribution of Sculptor Group
dIs in Figure 9.  The first is the average displacement 
of the Sculptor Group dIs on the vertical axis relative to the Local 
Group dIs.  The Sculptor Group dIs are very similar in gas content to
the HI-rich low surface brightness galaxies studied by van Zee et al.\ 
(1997a,b).  This is especially noteworthy since
the van Zee et al.\ (1997a) sample was selected
specifically from galaxies with the largest HI gas contents. 
While galaxies of this type appear to be common in the Sculptor Group,
clearly it is rare to find Local Group dwarf galaxies
with such large HI gas contents.  
Of the Local Group sample, only NGC~3109 has 
a comparably large HI gas content. 

Second, of the Sculptor Group dIs observed so far, there appears to
be a complete absence of the low metallicity, low gas content galaxies
which are common in the Local Group (e.g., Leo A, Sag DIG, and the other
galaxies in the lower left corner of Figure 9).  
Although we have not yet obtained spectra for all of the lowest 
luminosity (and therefore lowest metallicity) dIs in the Sculptor 
Group, we have not applied a bias for selecting against these galaxies. 
At this point, the relative lack of low metallicity, low gas content 
galaxies in the Sculptor Group appears to be real.  Note that there are
different ways in which to produce low metallicity, low gas content 
galaxies.  Although a favored method is loss of metals, loss of
HI gas can produce the same result.  If the low gas contents of these 
galaxies are attributable to one of these other processes
besides galactic winds (e.g., ram pressure stripping, Lin \& Faber
1983, or ``tidal stirring'', Mayer et al.\ 2001a,b) then
this lack of low metallicity, low gas content dIs in the 
Sculptor Group could be a result of the lower density 
environment.  

Although the Sculptor Group dIs and the LSB dIs lie close to the
higher yield simple closed box model in Figure 9, the effective
yields for many of them do leave some room for some deviation
from the simple closed box model as either selective
loss of metals or stripping of HI.  Also note the following shortcoming 
of this simple diagram with regard to the assumption that M(HI)/L(B)
gives a close approximation to gas mass fraction.  In the case of
NGC~625,  low value of M(HI)/L(B) is due to its enhanced
luminosity because of the recent burst of star formation, and it lies
right next to the Pegasus dI with a low M(HI)/L(B) due to a
truly low gas mass fraction.  It is very likely that NGC~625, with
an active starburst, has a much lower value of M/L than most of the 
other galaxies plotted in Figure 9. Thus it would be better 
represented by a higher yield model than its present position 
indicates.  Replacing L(B) with I band or K band luminosity would
likely reduce the impact of the recent star formation on this 
diagnostic diagram.  Surface photometry of these relatively low
surface brightness systems at longer wavelengths, while difficult,
is still highly desirable.
  
\section{Conclusions}

Using the CTIO 4-m, we have obtained optical spectroscopy of HII 
regions in  five dI galaxies in the nearby Sculptor Group.  From these
spectra we derive oxygen, nitrogen, and sulfur abundances.
Oxygen abundances are derived via three different methods (the ``direct'' method,
the empirical method guided by photoionization modeling of McGaugh (1991), and
the purely empirical method of Pilyugin (2000)) and are compared. 
Significant systematic differences are found between the three methods, 
and we suggest
that a recalibration of the empirical abundance scale is required.
Until differences between these three methods are better understood, the
issue of the degree of uniformity of the ISM abundances in a dwarf galaxy
cannot be properly addressed.

The N/O ratio in ESO~473-G24, which is a very metal poor galaxy,
is found to be in the normal range for dwarf irregular galaxies, but 
elevated when compared to
the very narrow range found by IT99 for low metallicity BCGs.
This shows that not all galaxies with 12 $+$ log (O/H) $\le$ 7.6
have identical elemental abundance ratios, and this
implies that the Izotov \& Thuan scenario for low metallicity
galaxies is not universal.

Measurements of the HII regions in NGC~625 yield
log (N/O) $\approx$ $-$1.25.  This relatively high N/O ratio may be
indicative of a long quiescent period prior to the recent
active burst of star formation.  Altogether, the observations presented
here and in the literature can be seen to support the scenario
of N produced by intermediate mass stars.  The large scatter
in N/O at a given O/H is still consistent with contamination
by O followed by later contamination by N.  This is in agreement
with what is observed in the high redshift damped Ly$\alpha$ systems
(see Pettini et al.\ 2002 and references therein) and resolves the
conflict between high and low redshift observations discussed
by Pilyugin (1999).

The oxygen abundances in the Sculptor Group dIs are in good agreement 
with the relationship between metallicity and luminosity observed in 
the Local Group dIs.  Taken together the observations show a better
relationship between metallicity and luminosity than between 
metallicity and galaxy central surface brightness.  

  The gas contents of the Sculptor Group dIs are large compared to
the Local Group dIs.  Many of the Sculptor Group dIs lie close to the
evolutionary path predicted by a simple closed box model with an
effective oxygen yield that is close to the theoretically favored values.
It is possible that some of the Sculptor Group dwarfs are evolving nearly
as closed systems.  On the other hand, if the abundances in their 
extended HI disks are lower than in the HII regions, the appropriate value
of the effective yield would be much smaller.
The higher gas contents, lower average star formation rates,
and closer resemblance to closed box evolution for Sculptor Group dIs
relative to Local Group dIs could all be
indicative of evolution in a relatively lower density environment.

\acknowledgments

We are grateful for the help of CTIO staff members Manuel Hernandez
and Ricardo Venegas.
We wish to thank G.\ Bothun, M.\ Edmunds, D.\ Garnett, R.\ Henry,
R.\ Kennicutt, E.\ Tolstoy, and L. van Zee for many helpful conversations. 
John Cannon, Herny Lee, and Liese van Zee proofread and provided valuable 
comments on this manuscript.
We also thank the referee for a prompt and careful reading of the
manuscript and their valuable quibbles.
This research has made use of the NASA/IPAC Extragalactic Database 
(NED) which is operated by the Jet Propulsion Laboratory, California
Institute of Technology, under contract with the National Aeronautics 
and Space Administration.
This research has made use of NASA's Astrophysics Data System
Abstract Service. 
EDS acknowledges partial support from a NASA LTSARP grant No. NAG5-9221 
and the University of Minnesota.  
BWM is supported by the Gemini Observatory, which is operated by the
Association of Universities for Research in Astronomy, Inc., on behalf
of the international Gemini partnership of Argentina, Australia, Brazil,
Canada, Chile, the United Kingdom, and the United States of America.

\clearpage


\clearpage

\begin{deluxetable}{lcccc}
\footnotesize
\tablenum{1a}
\tablewidth{0pt}
\tablecaption{Corrected Relative Emission Line Fluxes and Errors \label{tbl-1a}}
\tablehead{
\colhead{Wavelength} & \colhead{f($\lambda$)}     &
\colhead{ESO~347-G17 \#5}   &
\colhead{ESO~347-G17 \#10}   &
\colhead{ESO~348-G09 \#3}   \\
\colhead{} & \colhead{}     &
\colhead{I($\lambda$)/I(H$\beta$)}   &
\colhead{I($\lambda$)/I(H$\beta$)}   &
\colhead{I($\lambda$)/I(H$\beta$)}
}
\startdata
3727 [O~II]    &  0.267 & 4.204 $\pm$ 0.131 & 3.079 $\pm$ 0.151 & 2.941 $\pm$ 0.159 \\
3868 [Ne~III]  &  0.240 & 0.404 $\pm$ 0.019 & 0.359 $\pm$ 0.034 &  ... \\
3889 H8 + He I &  0.237 & 0.271 $\pm$ 0.015 & 0.332 $\pm$ 0.030 &  ... \\
3969 [Ne~III] + H7&0.221& 0.269 $\pm$ 0.013 & ...               &  ... \\
4101 H$\delta$ &  0.193 & 0.242 $\pm$ 0.012 & 0.258 $\pm$ 0.023 & 0.262 $\pm$ 0.027 \\
4340 H$\gamma$ &  0.137 & 0.488 $\pm$ 0.015 & 0.463 $\pm$ 0.023 & 0.459 $\pm$ 0.027 \\
4363 [O~III]   &  0.137 & 0.061 $\pm$ 0.006 & ...               & ...  \\
4471 He I      &  0.109 & 0.030 $\pm$ 0.005 & ...               & ...  \\
4861 H$\beta$  &  0.000 & 1.000 $\pm$ 0.022 & 1.000 $\pm$ 0.026 & 1.000 $\pm$ 0.029 \\
4959 [O~III]   & -0.022 & 1.018 $\pm$ 0.022 & 1.587 $\pm$ 0.037 & 0.428 $\pm$ 0.017 \\
5007 [O~III]   & -0.022 & 3.086 $\pm$ 0.063 & 3.790 $\pm$ 0.081 & 1.093 $\pm$ 0.030 \\
5876 He I      & -0.225 & 0.064 $\pm$ 0.003 & 0.073 $\pm$ 0.007 & ...  \\
6300 [O~I]     & -0.292 & 0.046 $\pm$ 0.003 & ...               & ...  \\
6312 [S~III]   & -0.292 & 0.019 $\pm$ 0.002 & ...               & ...  \\
6363 [O~I]     & -0.301 & 0.016 $\pm$ 0.002 & ...               & ...  \\
6548 [N~II]    & -0.329 & 0.035 $\pm$ 0.003 & 0.029 $\pm$ 0.005 & ...  \\
6563 H$\alpha$ & -0.329 & 2.796 $\pm$ 0.091 & 2.809 $\pm$ 0.141 & 2.809 $\pm$ 0.148 \\
6584 [N~II]    & -0.329 & 0.116 $\pm$ 0.005 & 0.103 $\pm$ 0.008 & 0.173 $\pm$ 0.012 \\
6678 He I      & -0.346 & 0.026 $\pm$ 0.002 & ...               & ...  \\
6717 [S~II]    & -0.351 & 0.355 $\pm$ 0.013 & 0.207 $\pm$ 0.013 & 0.295 $\pm$ 0.019 \\
6731 [S~II]    & -0.351 & 0.243 $\pm$ 0.009 & 0.205 $\pm$ 0.013 & 0.186 $\pm$ 0.013 \\
7065 He I      & -0.396 & 0.022 $\pm$ 0.002 & ...               & ...  \\
                                      &                          \\
F(H$\beta$) (erg cm $^{-2}$ s$^{-1})$ & ... & 3.4 $\times$ 10$^{-15}$ &
1.0 $\times$ 10$^{-15}$ &  4.2 $\times$ 10$^{-16}$ \\
C(H$\beta$)          & ... & 0.29 $\pm$ 0.03 & 0.26 $\pm$ 0.06  & 0.16 $\pm$ 0.06 \\
EW(H$\beta$) (\AA )  & ... & 30   $\pm$   1  & 15   $\pm$ 1     & 26   $\pm$ 1    \\
EW(H abs) (\AA )     & ... & 0.3  $\pm$ 0.3  & 0.8  $\pm$ 0.4   & 1.5  $\pm$ 0.6  \\
\enddata

\end{deluxetable}

\clearpage

\begin{deluxetable}{lcccc}
\scriptsize
\tablenum{1b}
\tablewidth{0pt}
\tablecaption{Corrected Relative Emission Line Fluxes and Errors \label{tbl-1b}}
\tablehead{
\colhead{Wavelength} & \colhead{f($\lambda$)}     &
\colhead{ESO~471-G06 \#2}   &
\colhead{ESO~473-G24 \#2}   &
\colhead{ESO~473-G24 \#4}   \\
\colhead{} & \colhead{}     &
\colhead{I($\lambda$)/I(H$\beta$)}   &
\colhead{I($\lambda$)/I(H$\beta$)}   &
\colhead{I($\lambda$)/I(H$\beta$)}
}
\startdata
3727 [O~II]    &  0.267 & 2.072 $\pm$ 0.075 & 1.109 $\pm$ 0.044 & 1.138 $\pm$ 0.072 \\
3798 H10       &  0.253 & \nodata           & \nodata           & \nodata \\
3835 H9        &  0.248 & 0.072 $\pm$ 0.006 & 0.061 $\pm$ 0.006 &  ... \\
3868 [Ne~III]  &  0.240 & 0.225 $\pm$ 0.010 & 0.210 $\pm$ 0.011 & 0.184 $\pm$ 0.023 \\
3889 H8 + He I &  0.237 & 0.190 $\pm$ 0.008 & 0.202 $\pm$ 0.010 & 0.277 $\pm$ 0.026 \\
3969 [Ne~III] + H7&0.221& 0.189 $\pm$ 0.008 & 0.201 $\pm$ 0.009 & 0.147 $\pm$ 0.019 \\
4101 H$\delta$ &  0.193 & 0.253 $\pm$ 0.009 & 0.268 $\pm$ 0.010 & 0.272 $\pm$ 0.021 \\
4340 H$\gamma$ &  0.137 & 0.469 $\pm$ 0.013 & 0.471 $\pm$ 0.014 & 0.441 $\pm$ 0.021 \\
4363 [O~III]   &  0.137 & 0.050 $\pm$ 0.003 & 0.058 $\pm$ 0.004 & ...  \\
4471 He I      &  0.109 & 0.032 $\pm$ 0.003 & 0.036 $\pm$ 0.003 & 0.050 $\pm$ 0.011 \\
4861 H$\beta$  &  0.000 & 1.000 $\pm$ 0.021 & 1.000 $\pm$ 0.021 & 1.000 $\pm$ 0.026 \\
4959 [O~III]   & -0.022 & 0.848 $\pm$ 0.018 & 0.769 $\pm$ 0.016 & 0.566 $\pm$ 0.017 \\
5007 [O~III]   & -0.022 & 2.654 $\pm$ 0.054 & 2.291 $\pm$ 0.047 & 1.838 $\pm$ 0.043 \\
5876 He I      & -0.225 & 0.096 $\pm$ 0.003 & 0.093 $\pm$ 0.003 & 0.082 $\pm$ 0.007 \\
6300 [O~I]     & -0.292 & 0.025 $\pm$ 0.001 & 0.010 $\pm$ 0.001 & ...  \\
6312 [S~III]   & -0.292 & 0.015 $\pm$ 0.001 & 0.012 $\pm$ 0.001 & 0.025 $\pm$ 0.005 \\
6363 [O~I]     & -0.301 & 0.010 $\pm$ 0.001 & ...               & ...  \\
6548 [N~II]    & -0.329 & 0.032 $\pm$ 0.002 & 0.019 $\pm$ 0.001 & ...  \\
6563 H$\alpha$ & -0.329 & 2.796 $\pm$ 0.113 & 2.741 $\pm$ 0.113 & 2.771 $\pm$ 0.145 \\
6584 [N~II]    & -0.329 & 0.098 $\pm$ 0.004 & 0.046 $\pm$ 0.002 & 0.032 $\pm$ 0.005 \\
6678 He I      & -0.346 & 0.026 $\pm$ 0.001 & 0.028 $\pm$ 0.002 & ...  \\
6717 [S~II]    & -0.351 & 0.181 $\pm$ 0.008 & 0.070 $\pm$ 0.003 & 0.095 $\pm$ 0.007 \\
6731 [S~II]    & -0.351 & 0.126 $\pm$ 0.006 & 0.055 $\pm$ 0.003 & 0.050 $\pm$ 0.005 \\
7065 He I      & -0.396 & 0.018 $\pm$ 0.001 & 0.028 $\pm$ 0.002 & 0.016 $\pm$ 0.004 \\
                                      &                          \\
F(H$\beta$) (erg cm $^{-2}$ s$^{-1})$ & ... & 5.7 $\times$ 10$^{-15}$ &
2.3 $\times$ 10$^{-15}$ &  4.2 $\times$ 10$^{-16}$ \\
C(H$\beta$)          & ... & 0.11  $\pm$ 0.05 & 0.19  $\pm$ 0.05  & 0.14 $\pm$ 0.06 \\
EW(H$\beta$) (\AA )  & ... & 74   $\pm$   3  & 133  $\pm$ 6     & 54   $\pm$ 3    \\
EW(H abs) (\AA )     & ... & 1.3  $\pm$ 1.0  & 0.0  $\pm$ 1.6   & 2.5  $\pm$ 1.3  \\
\enddata

\end{deluxetable}
\clearpage

\begin{deluxetable}{lccccc}
\tabletypesize{\footnotesize}
\tablenum{1c}
\tablewidth{0pt}
\tablecaption{Corrected Relative Emission Line Fluxes and Errors \label{tbl-1c}}
\tablehead{
\colhead{Wavelength} & \colhead{f($\lambda$)}     &
\colhead{NGC~625  \#5}   &
\colhead{NGC~625  \#9}   &
\colhead{NGC~625  \#18}  &
\colhead{NGC~625  \#21} \\
\colhead{} & \colhead{}     &
\colhead{I($\lambda$)/I(H$\beta$)}   &
\colhead{I($\lambda$)/I(H$\beta$)}   &
\colhead{I($\lambda$)/I(H$\beta$)}   &
\colhead{I($\lambda$)/I(H$\beta$)}
}
\startdata
3727 [O~II]       &  0.267 & 1.7574 $\pm$ 0.0585 & 2.3144 $\pm$ 0.0814 & 3.379 $\pm$ 0.121 & 4.861 $\pm$ 0.476 \\
3750 H12          &  0.264 & 0.0300 $\pm$ 0.0013 & 0.0155 $\pm$ 0.0022 & ...               &   ... \\
3771 H11          &  0.259 & 0.0346 $\pm$ 0.0014 & 0.0305 $\pm$ 0.0024 & ...               &   ... \\
3798 H10          &  0.253 & 0.0489 $\pm$ 0.0018 & 0.0399 $\pm$ 0.0026 & ...               &   ... \\
3820 He I         &  0.250 & 0.0116 $\pm$ 0.0007 & ...                 & ...               &   ... \\
3835 H9           &  0.248 & 0.0778 $\pm$ 0.0026 & 0.0663 $\pm$ 0.0031 & 0.067 $\pm$ 0.005 &   ... \\
3868 [Ne~III]     &  0.240 & 0.3268 $\pm$ 0.0103 & 0.1996 $\pm$ 0.0071 & 0.179 $\pm$ 0.008 &   ... \\
3889 H8 + He I    &  0.237 & 0.1942 $\pm$ 0.0061 & 0.1841 $\pm$ 0.0065 & 0.185 $\pm$ 0.008 &   ... \\
3969 [Ne~III] + H7&  0.221 & 0.2653 $\pm$ 0.0080 & 0.1953 $\pm$ 0.0066 & 0.172 $\pm$ 0.007 &   ... \\
4026 HeI          &  0.208 & 0.0205 $\pm$ 0.0008 & 0.0161 $\pm$ 0.0017 & ...               &   ... \\
4068 [S~II]       &  0.200 & 0.0141 $\pm$ 0.0006 & 0.0165 $\pm$ 0.0016 & 0.029 $\pm$ 0.004 &   ... \\ 
4101 H$\delta$    &  0.193 & 0.2581 $\pm$ 0.0072 & 0.2619 $\pm$ 0.0079 & 0.255 $\pm$ 0.009 & 0.295 $\pm$ 0.041 \\
4340 H$\gamma$    &  0.137 & 0.4848 $\pm$ 0.0118 & 0.4718 $\pm$ 0.0120 & 0.469 $\pm$ 0.013 & 0.447 $\pm$ 0.038 \\
4363 [O~III]      &  0.137 & 0.0382 $\pm$ 0.0010 & 0.0219 $\pm$ 0.0013 & 0.034 $\pm$ 0.003 &   ...  \\
4387 He I         &  0.128 & 0.0050 $\pm$ 0.0003 & ...                 & ...               &   ...  \\
4471 He I         &  0.109 & 0.0457 $\pm$ 0.0011 & 0.0395 $\pm$ 0.0015 & 0.040 $\pm$ 0.003 &   ...  \\
4658 [Fe~III]     &  0.063 & 0.0048 $\pm$ 0.0003 & 0.0086 $\pm$ 0.0009 & ...               &   ... \\
4713 [Ar~IV] + He I& 0.050 & 0.0073 $\pm$ 0.0003 & 0.0057 $\pm$ 0.0009 & ...               &   ... \\
4861 H$\beta$     &  0.000 & 1.0000 $\pm$ 0.0200 & 1.0000 $\pm$ 0.0202 & 1.000 $\pm$ 0.020 & 1.000 $\pm$ 0.031 \\
4921 He I         & -0.001 & 0.0116 $\pm$ 0.0003 & 0.0079 $\pm$ 0.0008 & ...               &   ... \\
4959 [O~III]      & -0.022 & 1.5225 $\pm$ 0.0305 & 0.9890 $\pm$ 0.0200 & 0.866 $\pm$ 0.018 & 0.331 $\pm$ 0.022 \\
5007 [O~III]      & -0.022 & 4.5297 $\pm$ 0.0912 & 2.9520 $\pm$ 0.0596 & 2.547 $\pm$ 0.052 & 1.028 $\pm$ 0.032 \\
5199 [N~I]        & -0.068 & 0.0052 $\pm$ 0.0002 & 0.0075 $\pm$ 0.0007 & 0.012 $\pm$ 0.001 &   ... \\
5271 [Fe~III]     & -0.085 & 0.0017 $\pm$ 0.0002 & ...                 & ...               &   ... \\
5518 [Cl~III]     & -0.166 & 0.0041 $\pm$ 0.0002 & 0.0041 $\pm$ 0.0006 & ...               &   ... \\
5538 [Cl~III]     & -0.169 & 0.0024 $\pm$ 0.0002 & 0.0027 $\pm$ 0.0006 & ...               &   ... \\
5876 He I         & -0.225 & 0.1164 $\pm$ 0.0035 & 0.1157 $\pm$ 0.0037 & 0.105 $\pm$ 0.004 & 0.084 $\pm$ 0.014 \\
6300 [O~I]        & -0.292 & 0.0201 $\pm$ 0.0007 & 0.0168 $\pm$ 0.0008 & 0.046 $\pm$ 0.002 &   ... \\
6312 [S~III]      & -0.292 & 0.0215 $\pm$ 0.0008 & 0.0201 $\pm$ 0.0009 & 0.017 $\pm$ 0.001 &   ... \\
6363 [O~I]        & -0.301 & 0.0064 $\pm$ 0.0003 & 0.0047 $\pm$ 0.0005 & 0.011 $\pm$ 0.001 &   ... \\
6548 [N~II]       & -0.329 & 0.0464 $\pm$ 0.0018 & 0.0706 $\pm$ 0.0029 & 0.084 $\pm$ 0.004 & 0.152 $\pm$ 0.021 \\
6563 H$\alpha$    & -0.329 & 2.8596 $\pm$ 0.1094 & 2.8447 $\pm$ 0.1152 & 2.818 $\pm$ 0.116 & 2.858 $\pm$ 0.333 \\
6584 [N~II]       & -0.329 & 0.1235 $\pm$ 0.0047 & 0.2039 $\pm$ 0.0083 & 0.262 $\pm$ 0.011 & 0.422 $\pm$ 0.050 \\
6678 He I         & -0.346 & 0.0356 $\pm$ 0.0014 & 0.0337 $\pm$ 0.0015 & 0.029 $\pm$ 0.002 &   ... \\
6717 [S~II]       & -0.351 & 0.1356 $\pm$ 0.0055 & 0.1966 $\pm$ 0.0084 & 0.341 $\pm$ 0.015 & 0.676 $\pm$ 0.085 \\
6731 [S~II]       & -0.351 & 0.1010 $\pm$ 0.0041 & 0.1464 $\pm$ 0.0063 & 0.240 $\pm$ 0.011 & 0.521 $\pm$ 0.066 \\
7065 He I         & -0.396 & 0.0272 $\pm$ 0.0012 & 0.0262 $\pm$ 0.0014 & 0.017 $\pm$ 0.001 &   ...  \\
                                      &                          \\
F(H$\beta$) (erg cm $^{-2}$ s$^{-1})$ & ... & 9.2 $\times$ 10$^{-14}$ &
2.0 $\times$ 10$^{-14}$ &  8.7 $\times$ 10$^{-15}$ & 4.7 $\times$ 10$^{-16}$ \\
C(H$\beta$)          & ... & 0.15 $\pm$ 0.04  & 0.01 $\pm$ 0.05  & 0.26 $\pm$ 0.05 & 0.32 $\pm$ 0.15 \\
EW(H$\beta$) (\AA )  & ... & 235  $\pm$   9   &  79  $\pm$ 3     & 37   $\pm$ 2    & 4.3  $\pm$ 0.3  \\
EW(H abs) (\AA )     & ... & 0.0  $\pm$ 2.4   & 0.5  $\pm$ 0.3   & 0.8  $\pm$ 0.5  & 0.8  $\pm$ 0.5  \\
\enddata
\end{deluxetable}

\clearpage

\begin{deluxetable}{lcccccc}
\tabletypesize{\footnotesize}
\rotate
\tablenum{2}
\tablewidth{0pt}
\tablecaption{Ionic and Total Abundances \label{tbl-2}}
\tablehead{
\colhead{Property} &
\colhead{ESO~347-G17 \#5} &
\colhead{ESO~471-G06 \#2} &
\colhead{ESO~473-G24 \#2} &
\colhead{NGC~625 \#5} &
\colhead{NGC~625 \#9} &
\colhead{NGC~625 \#18}
}
\startdata
T(O~III) (K)              & 15,170 $^{+840}_{-680}$ & 14,930 $^{+450}_{-400}$ & 17,080 $^{+620}_{-540}$  & 10,900 $^{+115}_{-109}$ & 10,460 $^{+213}_{-193}$ & 12,810 $^{+460}_{-395}$  \\
T(O~II) (K) (estimated)   & 13,370 $\pm$ 690  & 13,610 $\pm$ 500 & 14,440 $\pm$ 500 & 11,640 $\pm$ 500 & 11,390 $\pm$ 500 & 12,650 $\pm$ 500  \\
T(S~III) (K) (estimated)  & 14,300 $\pm$ 720  & 14,090 $\pm$ 500 & 15,880 $\pm$ 540 & 10,740 $\pm$ 500 & 10,380 $\pm$ 500 & 12,330 $\pm$ 500  \\
n(S~II)                   & $\le$ 31 (1 $\sigma$)& $\le$ 61 (1 $\sigma$) & 156$^{+102}_{-86}$ & 75$^{+85}_{-72}$   & 75$^{+90}_{-75}$   & $\le$ 84 (1 $\sigma$)  \\
                          &                    \\
                          &                    \\
O$^+$/H$^+$ (x 10$^5$)    & 4.75 $\pm$ 0.77   & 2.40 $\pm$ 0.30  & 1.06 $\pm$ 0.12 & 3.52 $\pm$ 0.57  & 5.05 $\pm$ 0.85 & 5.00 $\pm$ 0.70  \\
O$^{++}$/H$^+$ (x 10$^5$) & 3.18 $\pm$ 0.37   & 2.82 $\pm$ 0.19  & 1.79 $\pm$ 0.13 & 11.9 $\pm$ 0.44  & 8.89 $\pm$ 0.57 & 4.12 $\pm$ 0.37  \\
O/H    (x 10$^5$)         & 7.93 $\pm$ 0.86   & 5.22 $\pm$ 0.35  & 2.84 $\pm$ 0.18 & 15.4 $\pm$ 0.72  & 13.9 $\pm$ 1.03 & 9.12 $\pm$ 0.79 \\
12 + log (O/H)            & 7.90 $\pm$ 0.09   & 7.72 $\pm$ 0.03  & 7.45 $\pm$ 0.03 & 8.19 $\pm$ 0.02  & 8.14 $\pm$ 0.03 & 7.96 $\pm$ 0.04 \\
                          &                    \\
N$^+$/O$^+$ (x 10$^2$)    & 2.28 $\pm$ 0.13   & 3.86 $\pm$ 0.22  & 3.68 $\pm$ 0.24  & 4.68 $\pm$ 0.27   & 5.68 $\pm$ 0.34 & 5.79 $\pm$ 0.34 \\
log (N/O)                 & -1.64 $\pm$ 0.02  & -1.41 $\pm$ 0.02 & -1.43 $\pm$ 0.03 & -1.33 $\pm$ 0.02  &-1.25 $\pm$ 0.03 &-1.24 $\pm$ 0.25 \\
N/H    (x 10$^6$)         & 1.81 $\pm$ 0.22   & 2.02 $\pm$ 0.18  & 1.05 $\pm$ 0.10  & 7.21 $\pm$ 0.53   & 7.92 $\pm$ 0.75 & 5.28 $\pm$ 0.55 \\
                          &                    \\
S$^+$/H$^+$ (x 10$^7$)    & 6.98 $\pm$ 0.77   & 3.64 $\pm$ 0.34  & 1.33 $\pm$ 0.13 & 3.83 $\pm$ 0.42   & 5.83 $\pm$ 0.67 & 7.91 $\pm$ 0.80 \\
S$^{++}$/H$^+$ (x 10$^7$) & 12.50 $\pm$ 2.36  & 10.40$\pm$ 1.38  & 5.85 $\pm$ 0.68 & 78.6 $\pm$ 7.31   & 41.8 $\pm$ 8.46 & 18.1 $\pm$ 2.86 \\
ICF                       & 1.15 $\pm$ 0.02   & 1.22 $\pm$ 0.02  & 1.27 $\pm$ 0.03 & 1.49 $\pm$ 0.05   & 1.28 $\pm$ 0.03 & 1.18 $\pm$ 0.02 \\
S/H    (x 10$^6$)         & 2.24 $\pm$ 0.25   & 1.71 $\pm$ 0.15  & 0.91 $\pm$ 0.82 & 6.37 $\pm$ 0.76   & 6.11 $\pm$ 0.86 & 3.06 $\pm$ 0.30 \\
12 + log (S/H)            & 6.35 $\pm$ 0.05   & 6.23 $\pm$ 0.04  & 5.96 $\pm$ 0.04 & 6.80 $\pm$ 0.05   & 6.79 $\pm$ 0.06 & 6.49 $\pm$ 0.04 \\
S/O (x 10$^2$)            & 2.82 $\pm$ 0.44   & 3.28 $\pm$ 0.36  & 3.21 $\pm$ 0.35 & 4.14 $\pm$ 0.53   & 4.38 $\pm$ 0.70 & 3.35 $\pm$ 0.44 \\
log (S/O)                 & -1.55 $\pm$ 0.06  & -1.48 $\pm$ 0.04 &-1.49 $\pm$ 0.05 &-1.38 $\pm$ 0.05   &-1.36 $\pm$ 0.06 &-1.47 $\pm$ 0.05 \\
                          &                    \\
\enddata

\end{deluxetable}

\clearpage

\begin{deluxetable}{lccccc}
\footnotesize
\tablenum{3}
\tablewidth{0pt}
\tablecaption{Comparison of Direct and Empirical Oxygen Abundances \label{tbl-3}}
\tablehead{
\colhead{HII Region} & \colhead{Direct}     &
\colhead{log(R$_{23}$)} & \colhead{log(O32)}     &
\colhead{M91} & \colhead{P00}}
\startdata
ESO~347-G17 \#5   & 7.90 $\pm$ 0.09 & 0.920 $\pm$ 0.007 & -0.010 $\pm$ 0.015 & 8.25 $\pm$ 0.05 & 8.22 $\pm$ 0.02 \\
ESO~347-G17 \#10  & \nodata         & 0.927 $\pm$ 0.008 &  0.242 $\pm$ 0.022 & 8.35 $\pm$ 0.10 & 8.04 $\pm$ 0.02 \\
ESO~348-G09 \#3   & \nodata         & 0.649 $\pm$ 0.015 & -0.287 $\pm$ 0.025 & 7.89 $\pm$ 0.03 & 8.11 $\pm$ 0.05 \\
ESO~471-G06 \#2   & 7.72 $\pm$ 0.03 & 0.746 $\pm$ 0.006 &  0.228 $\pm$ 0.017 & 7.83 $\pm$ 0.02 & 7.78 $\pm$ 0.02 \\
ESO~473-G24 \#2   & 7.45 $\pm$ 0.03 & 0.620 $\pm$ 0.005 &  0.441 $\pm$ 0.018 & 7.60 $\pm$ 0.02 & 7.48 $\pm$ 0.02 \\
ESO~473-G24 \#4   & \nodata         & 0.549 $\pm$ 0.009 &  0.325 $\pm$ 0.028 & 7.54 $\pm$ 0.02 & 7.44 $\pm$ 0.03 \\
NGC~625  \#5   & 8.19 $\pm$ 0.02 & 0.893 $\pm$ 0.003 &  0.537 $\pm$ 0.016 & 8.00 $\pm$ 0.02 & 7.84 $\pm$ 0.01 \\
NGC~625  \#9   & 8.14 $\pm$ 0.03 & 0.796 $\pm$ 0.006 &  0.231 $\pm$ 0.016 & 7.92 $\pm$ 0.02 & 7.85 $\pm$ 0.02 \\
NGC~625  \#18  & 7.96 $\pm$ 0.04 & 0.832 $\pm$ 0.008 &  0.010 $\pm$ 0.040 & 8.08 $\pm$ 0.03 & 8.08 $\pm$ 0.02 \\
NGC~625  \#21  & \nodata         & 0.794 $\pm$ 0.032 & -0.554 $\pm$ 0.042 & 8.31 $\pm$ 0.10 & 8.65 $\pm$ 0.10 \\
\enddata

\end{deluxetable}

\begin {figure}
\plotone{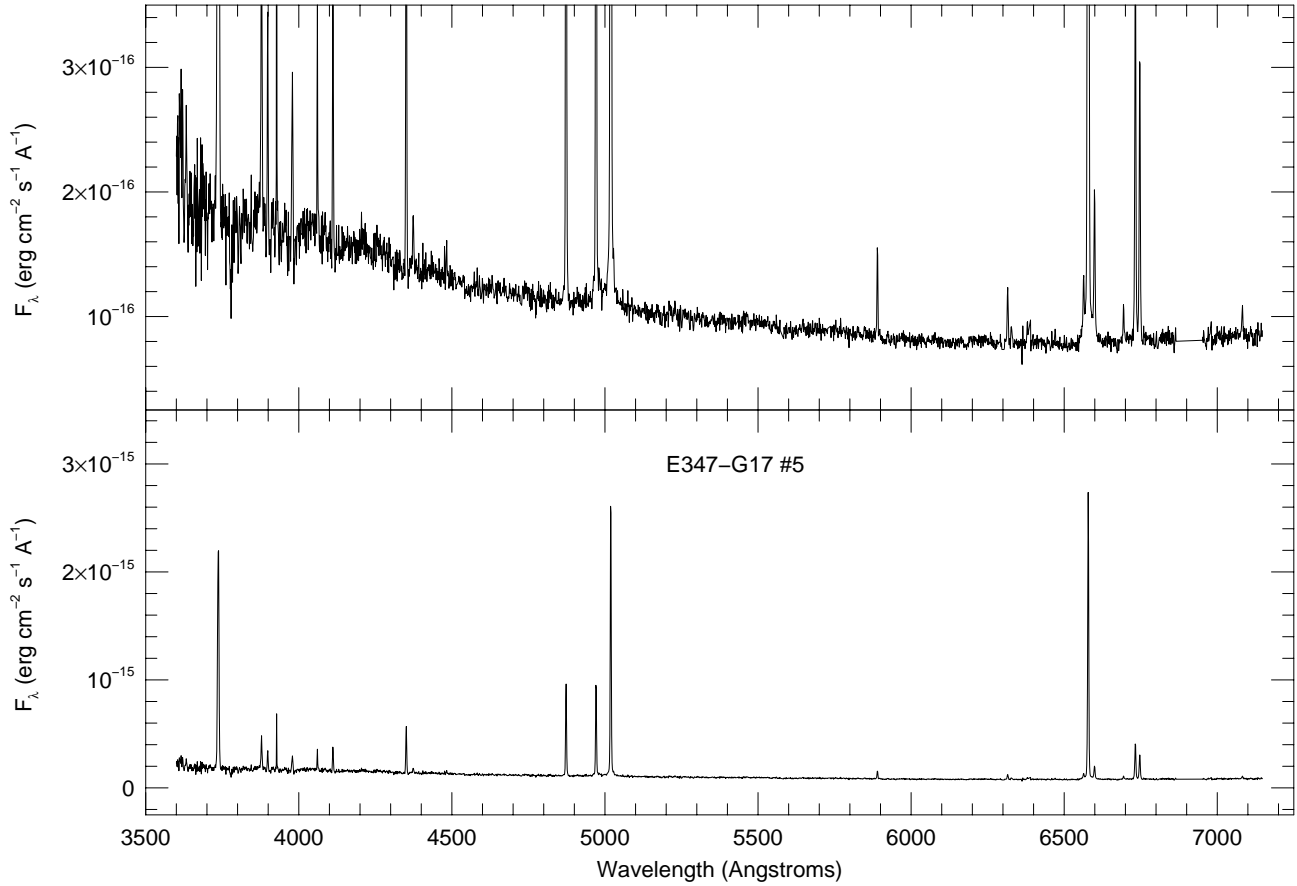}
\figcaption{Spectra of the brightest HII regions in each of the five
Sculptor Group dwarf irregular galaxies observed.  For four of the spectra,
the upper panel shows a greatly expanded scale in order to show the
faintest lines clearly.
\label{fig1}}
\end {figure}

\begin {figure}
\plotone{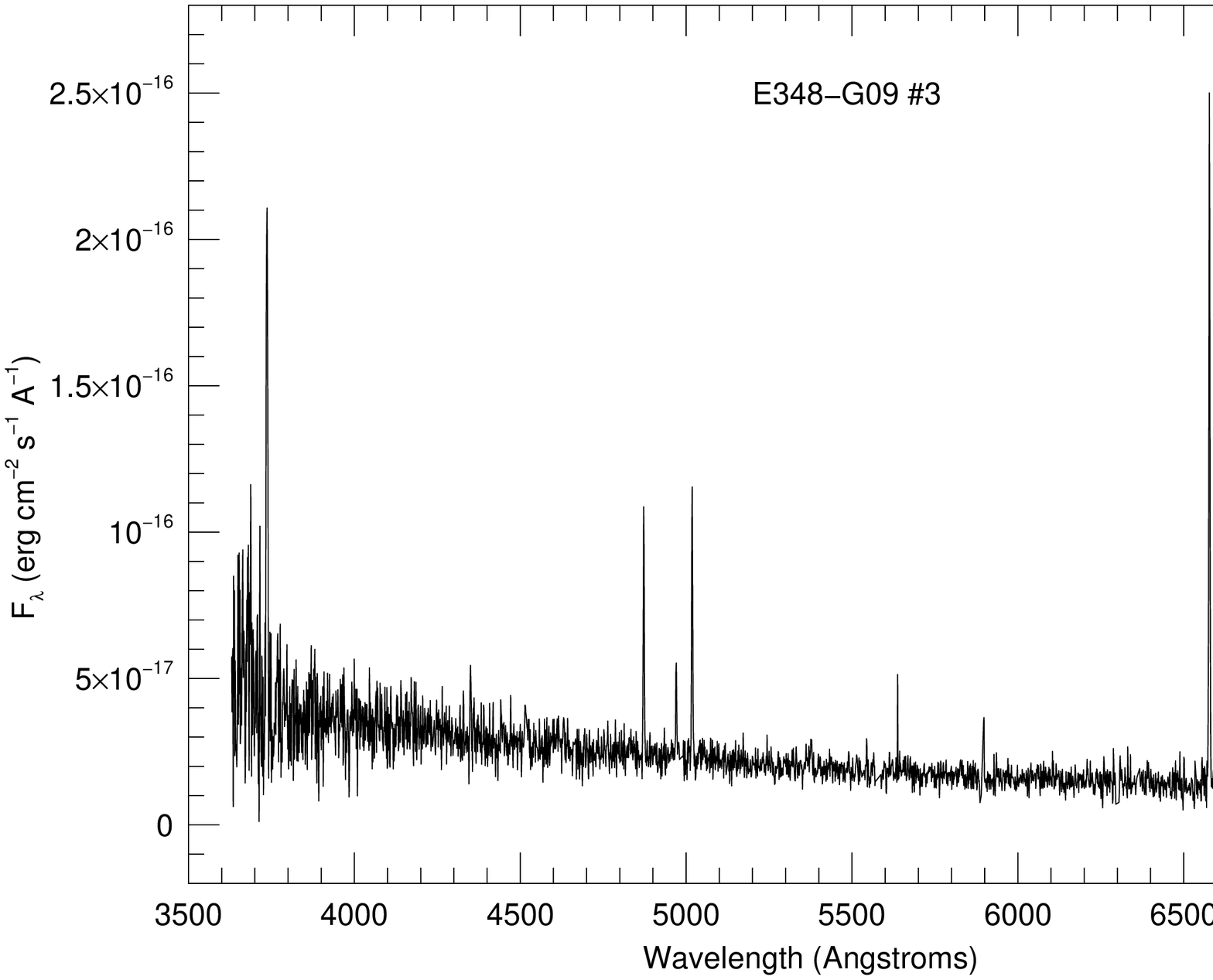}
\end {figure}

\begin {figure}
\plotone{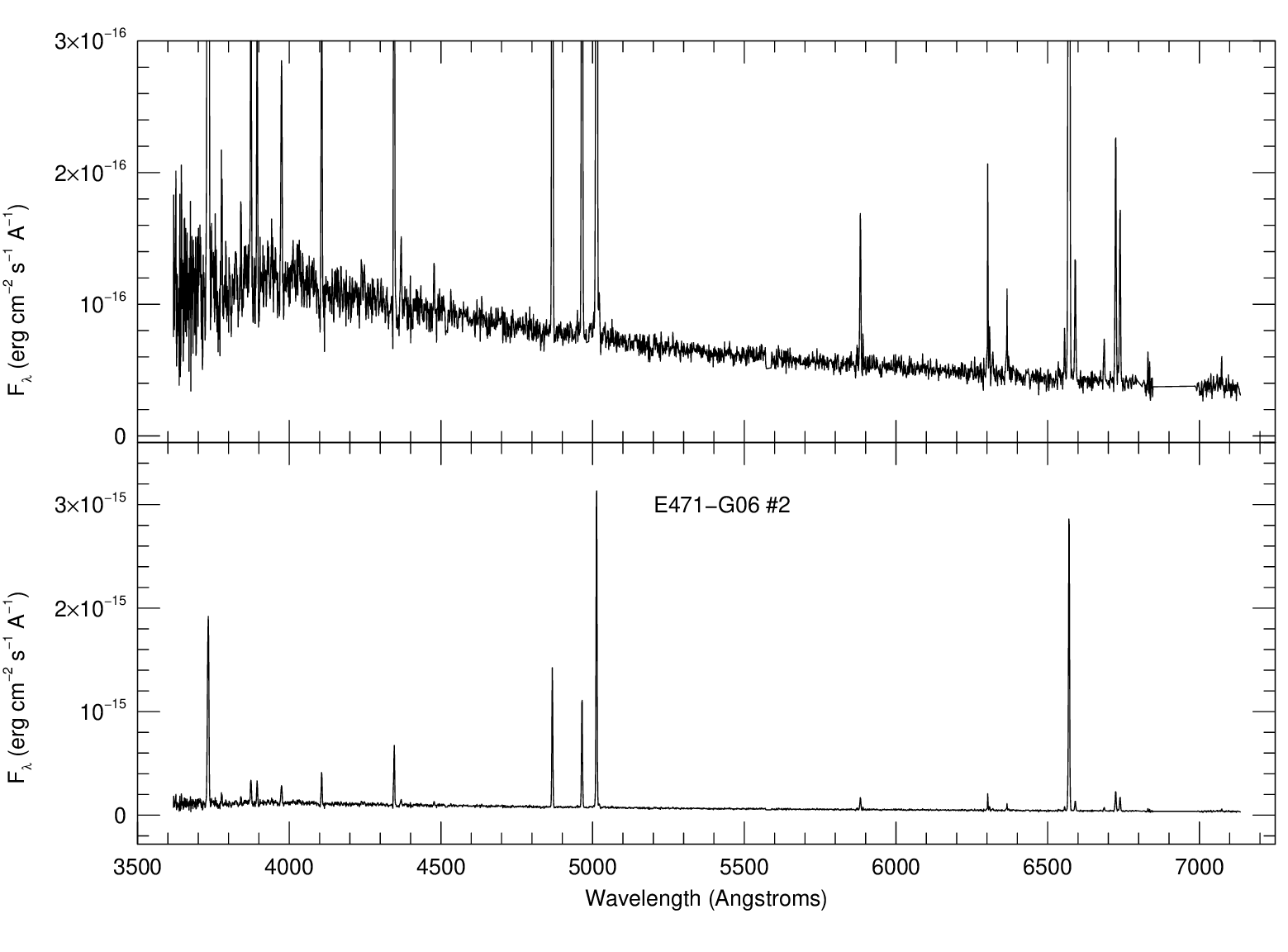}
\end {figure}

\begin {figure}
\plotone{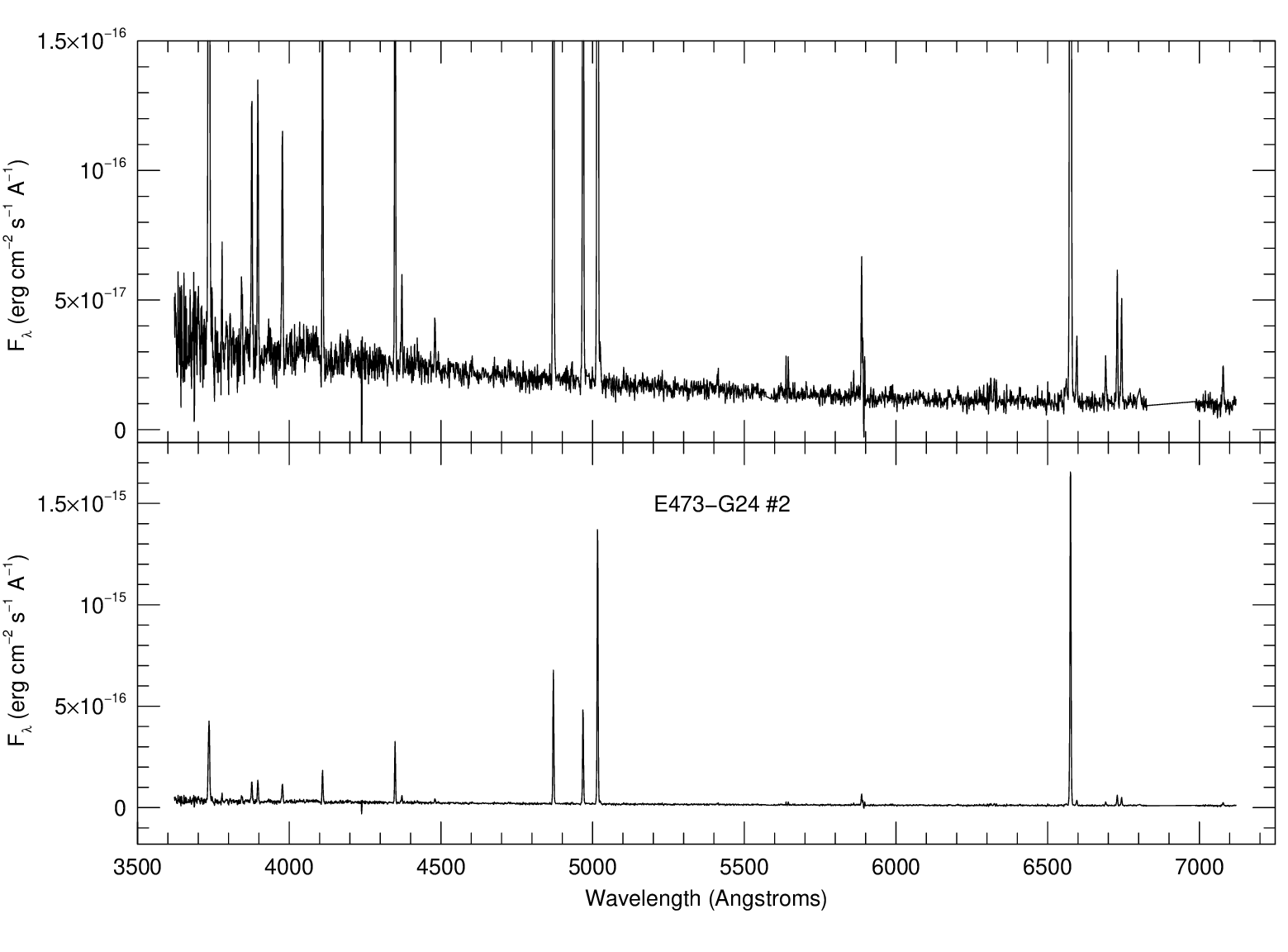}
\end {figure}

\begin {figure}
\plotone{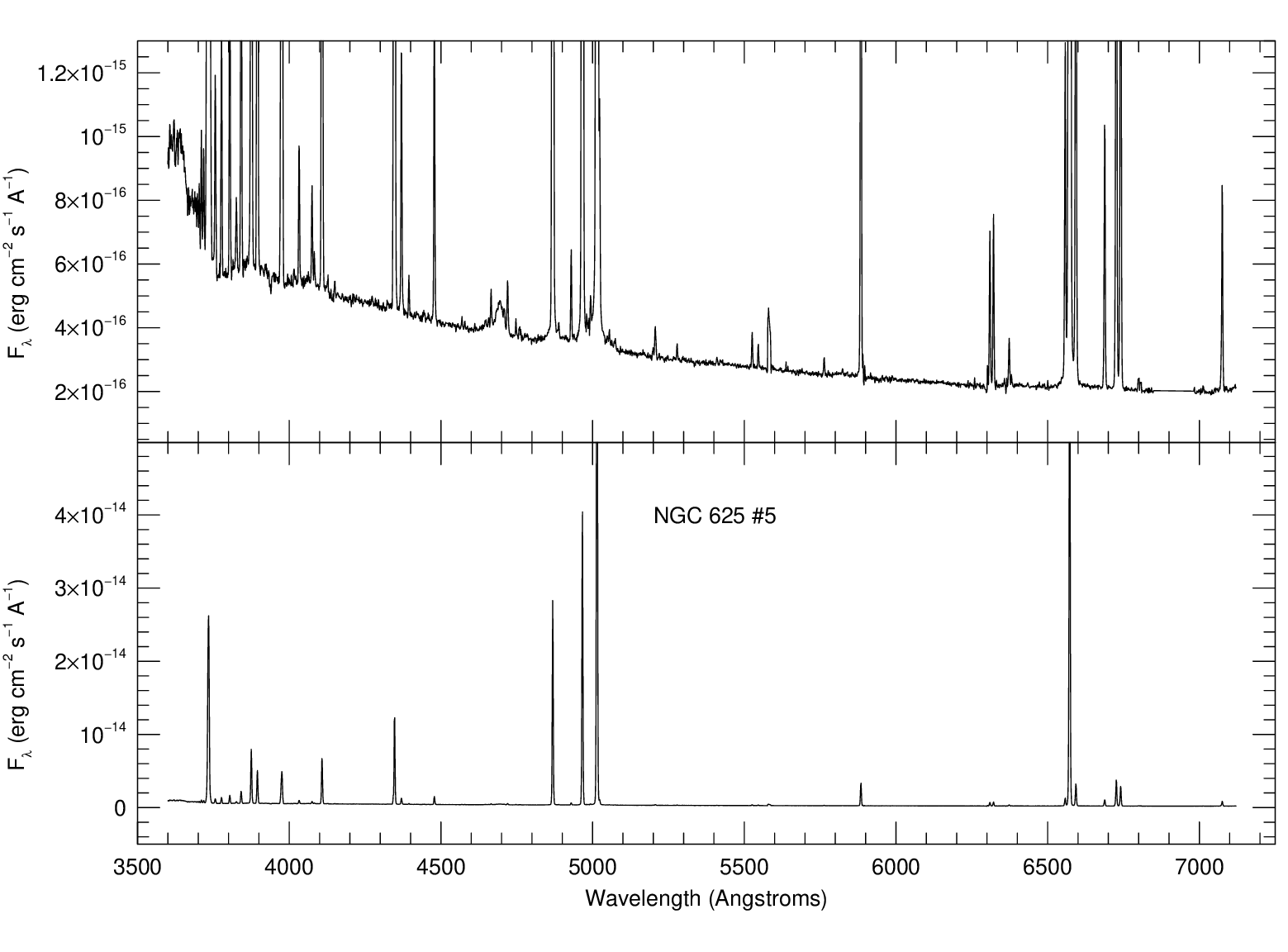}
\end {figure}

\clearpage 
\begin {figure}
\plotone{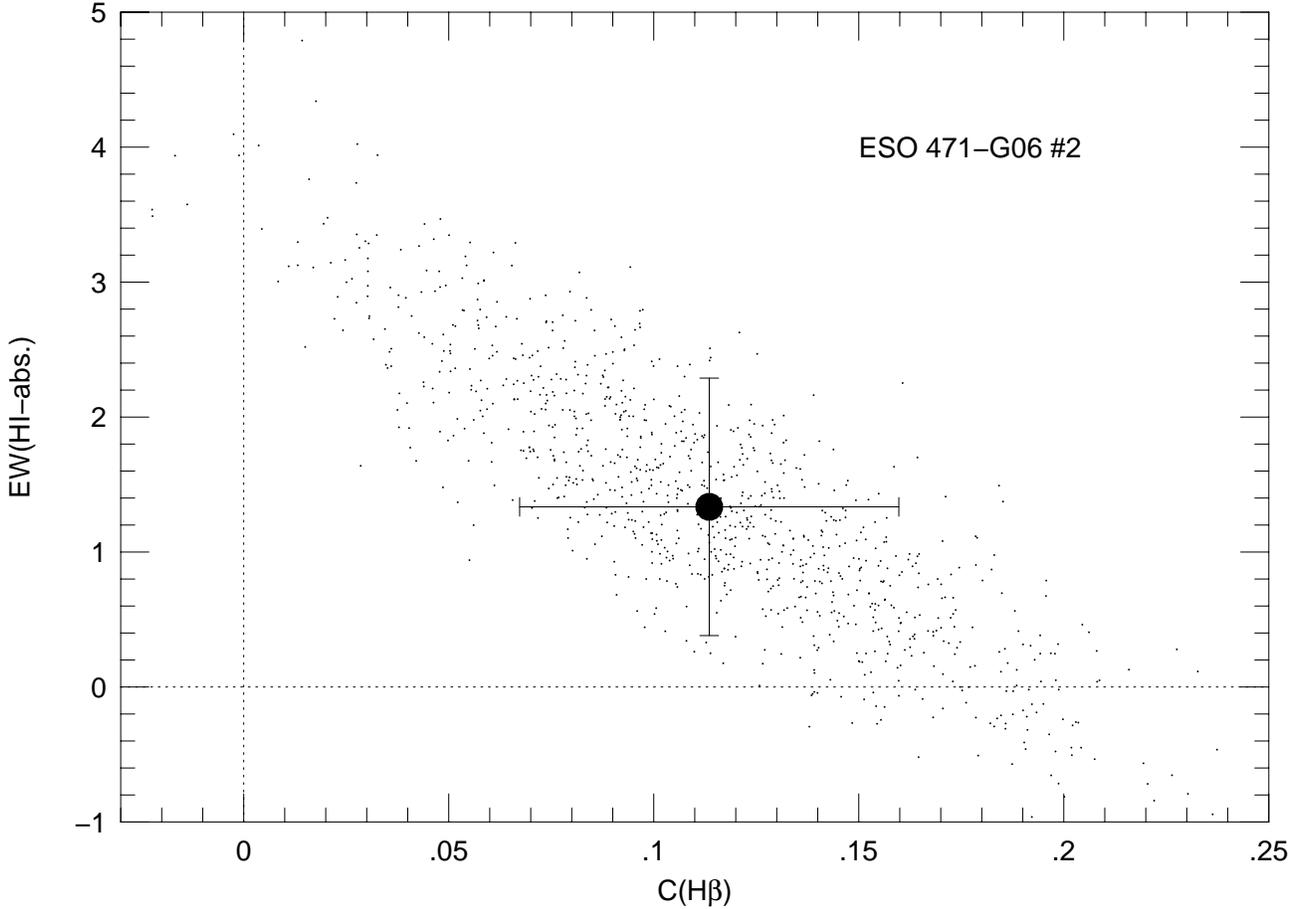}
\figcaption {
The results of Monte Carlo simulations of solutions for the reddening, 
CH$\beta$, and underlying absorption, EW(HI-abs.) from hydrogen Balmer 
emission line ratios for ESO~471-G05 \# 2.
Each small point is the solution derived from a different
realization of the same input spectrum and errors as given in Table 1.
The large filled  
point with error bars shows the mean result with 1$\sigma$ errors
derived from the dispersion in the solutions.  
Note that the covariance of the two parameters leads to error 
ellipses.  Because of the covariance, the error
bars appropriate to these solutions are about twice the
size of the errors inferred from a single $\chi^2$ minimization.
\label{fig2}}
\end{figure}

\clearpage 

\begin {figure}
\plotone{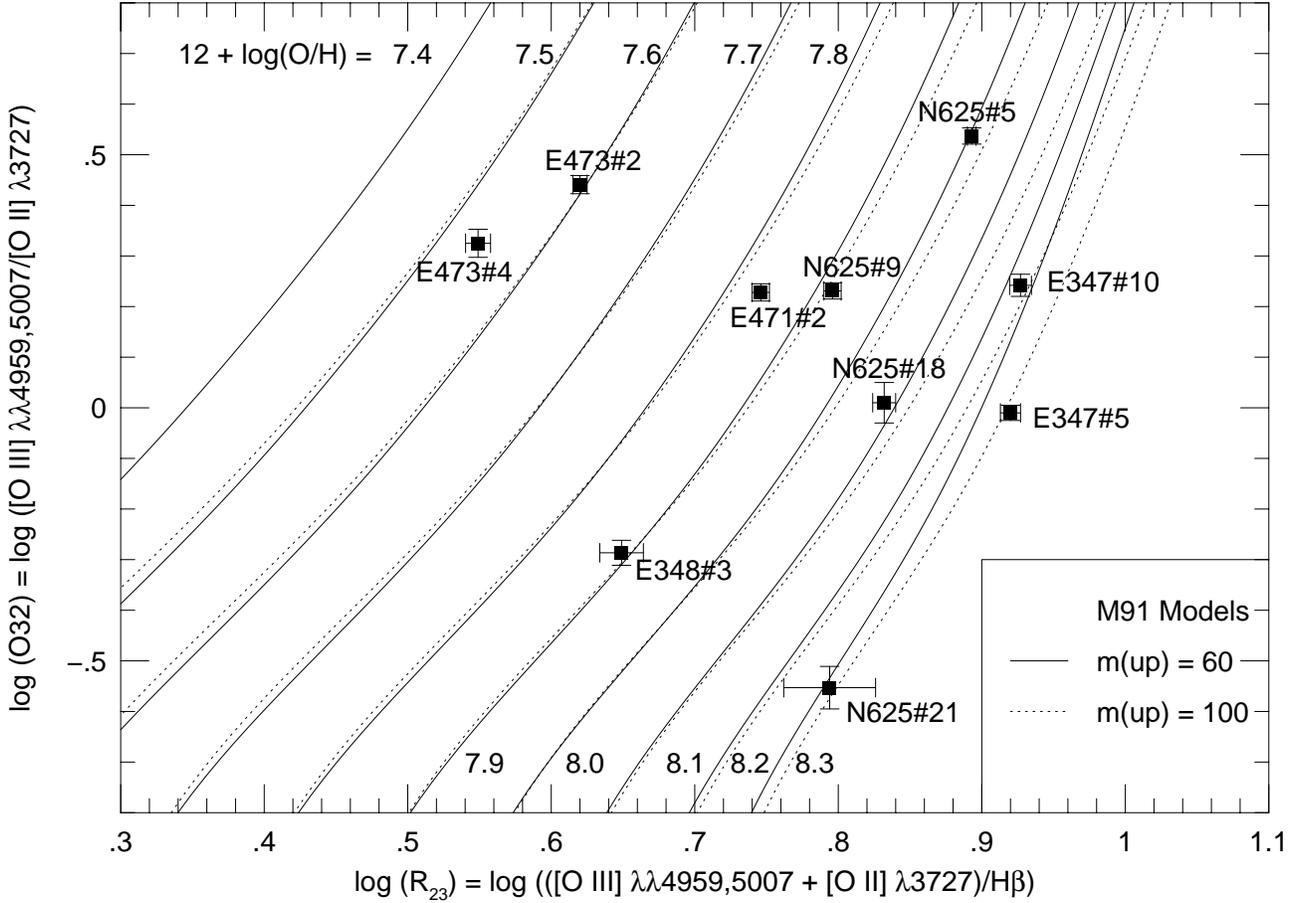}
\figcaption {Diagnostic diagram of oxygen emission line ratios
comparing observed quantities with the photoionization model grid of
McGaugh (1991). 
The two different sets of McGaugh (1991) models correspond to upper
limits on the IMF of 60 and 100 M$_\odot$, and emphasizes his point
that there is not a great difference between the two. 
The results for all ten spectra from Tables 1a - 1c are
presented here.  Oxygen abundances derived from the McGaugh models
are given in Table 3.  See text and next two figures for a comparison of 
oxygen abundances.  
\label{fig3}}
\end{figure}

\clearpage 

\begin {figure}
\plotone{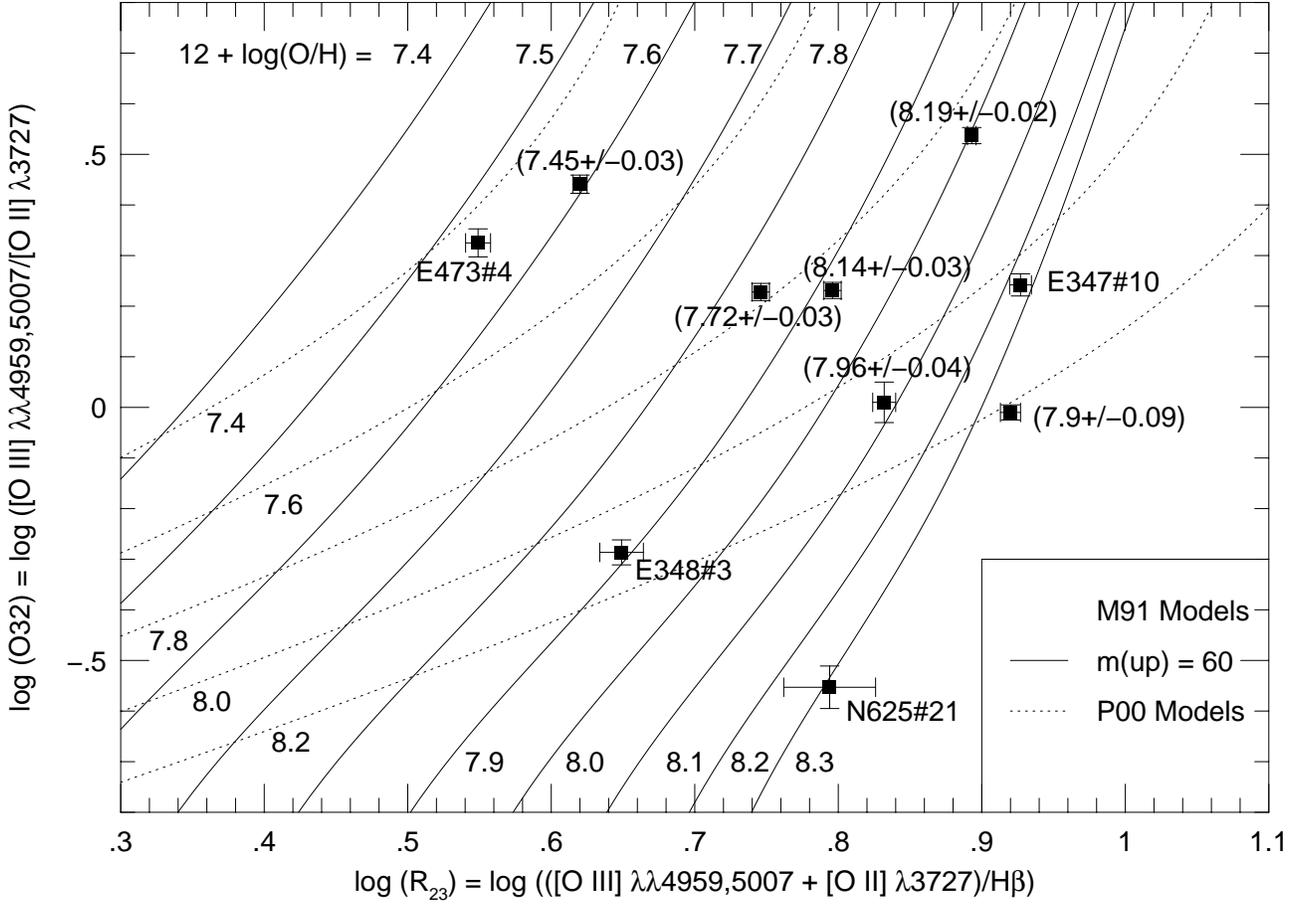}
\figcaption {Diagnostic diagram of oxygen emission line ratios as in
Figure 3, but comparing the observed quantities with both the photoionization 
models of McGaugh (1991) and the empirical abundances derived via the Pilyugin (2000)
method.  For those HII regions where [O~III] $\lambda$4363 has been observed, the 
result of the oxygen abundance derived via the direct method is given as a label.  
\label{fig4}}
\end{figure}

\clearpage

\begin {figure}
\plotone{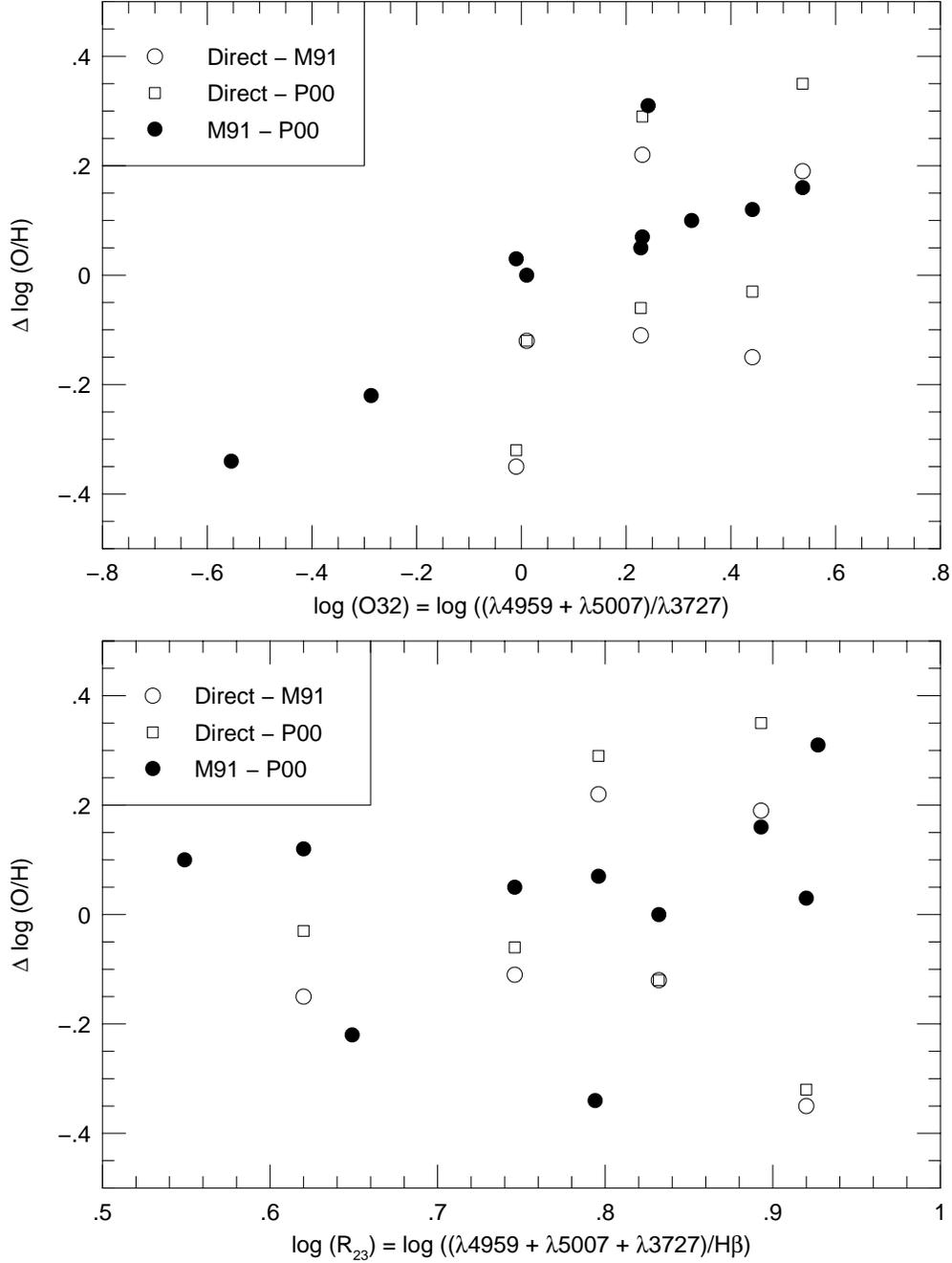}
\figcaption { Plots of the differences in oxygen abundances derived via three
different methods: ``Direct'' means that an electron temperature derived from the
[O~III] $\lambda$4363/($\lambda$4959 + $\lambda$5007) ratio has been used,
M91 means that the grids of McGaugh (1991) has been used, and P00 means that
the calibration of Pilyugin (2000) has been used.  The differences have been
plotted versus log (O32) and log (R$_{23}$).  In the plot of $\Delta$ log (O/H) versus
log (O32) there is a strong trend in the difference between the calibrations of
McGaugh (1991) and Pilyugin (2000).
\label{fig5}}
\end{figure}

\clearpage

\begin {figure}
\plotone{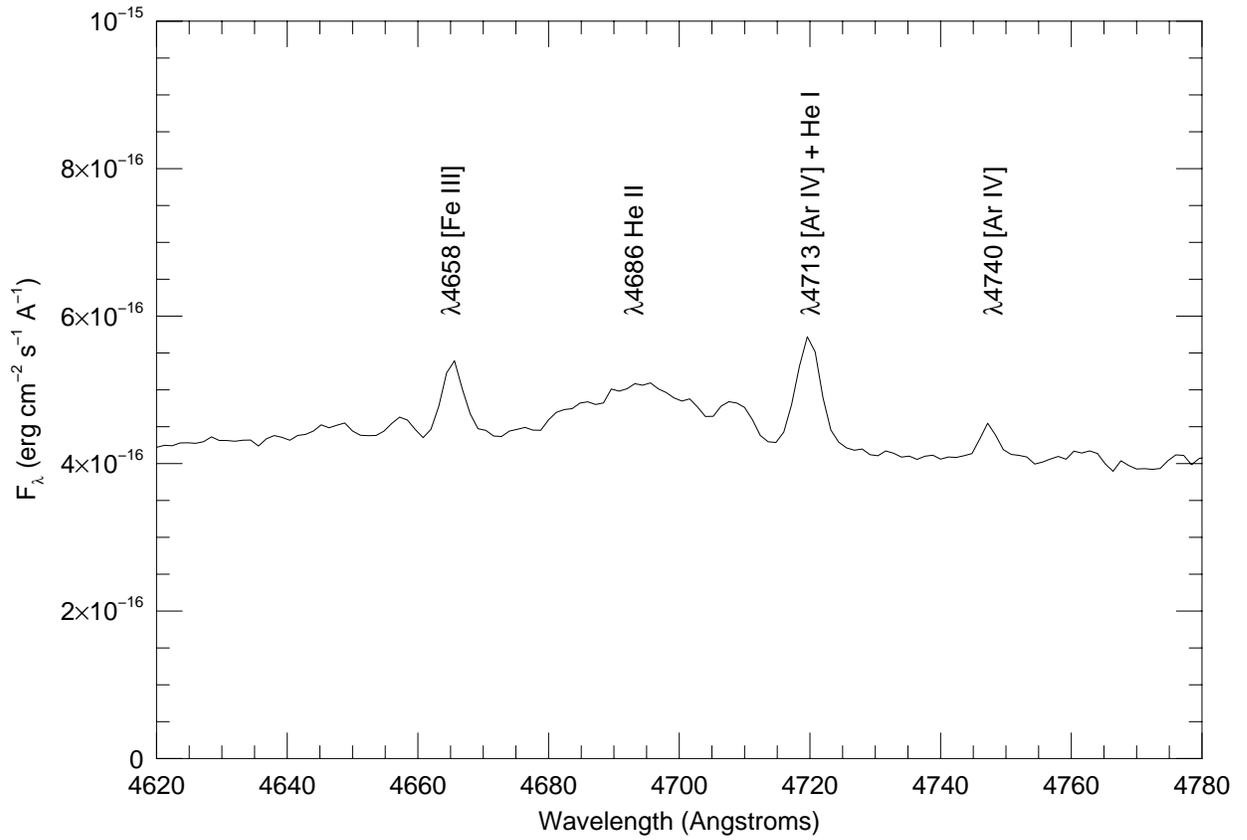}
\figcaption { Spectrum of NGC~625 \#5 in the region of $\sim$$\lambda$4700
where the broad emission of He II is observed.
\label{fig6}}
\end{figure}

\clearpage 

\begin {figure}
\plotone{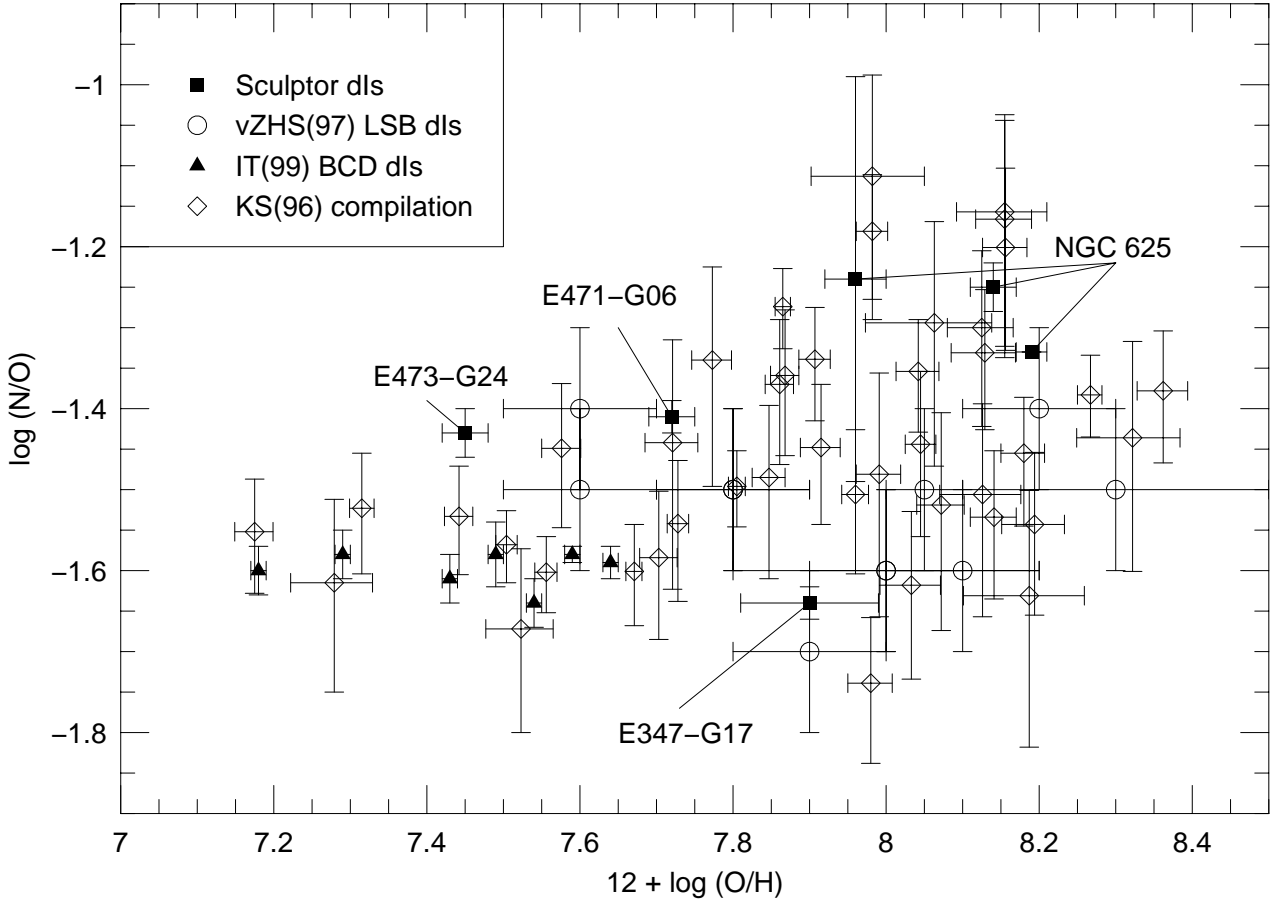}
\figcaption[j] {A comparison of the N/O and O/H in the Sculptor dwarf
irregular galaxies with other star forming dwarf galaxies from the
literature.  The collection of dwarf irregular galaxies and H~II galaxies
assembled by Kobulnicky \& Skillman (1996; see their Table 5 and Figure 15
for identification of individual points) are represented by open diamonds.
Only galaxies without WR emission features
and errors in log (N/O) less than 0.2 have been plotted.
The empty circular symbols represent data for low surface brightness dwarf
irregular galaxies from van Zee et al.\ 1997a. The filled
triangles represent the low metallicity blue compact dwarf galaxies
from Izotov \& Thuan (1999).  The four Sculptor dwarf irregular galaxies
with direct abundance measurements are shown with filled squares
(from the data presented in Table 2).
\label{fig7}}
\end{figure}

\clearpage 

\begin {figure}
\plotone{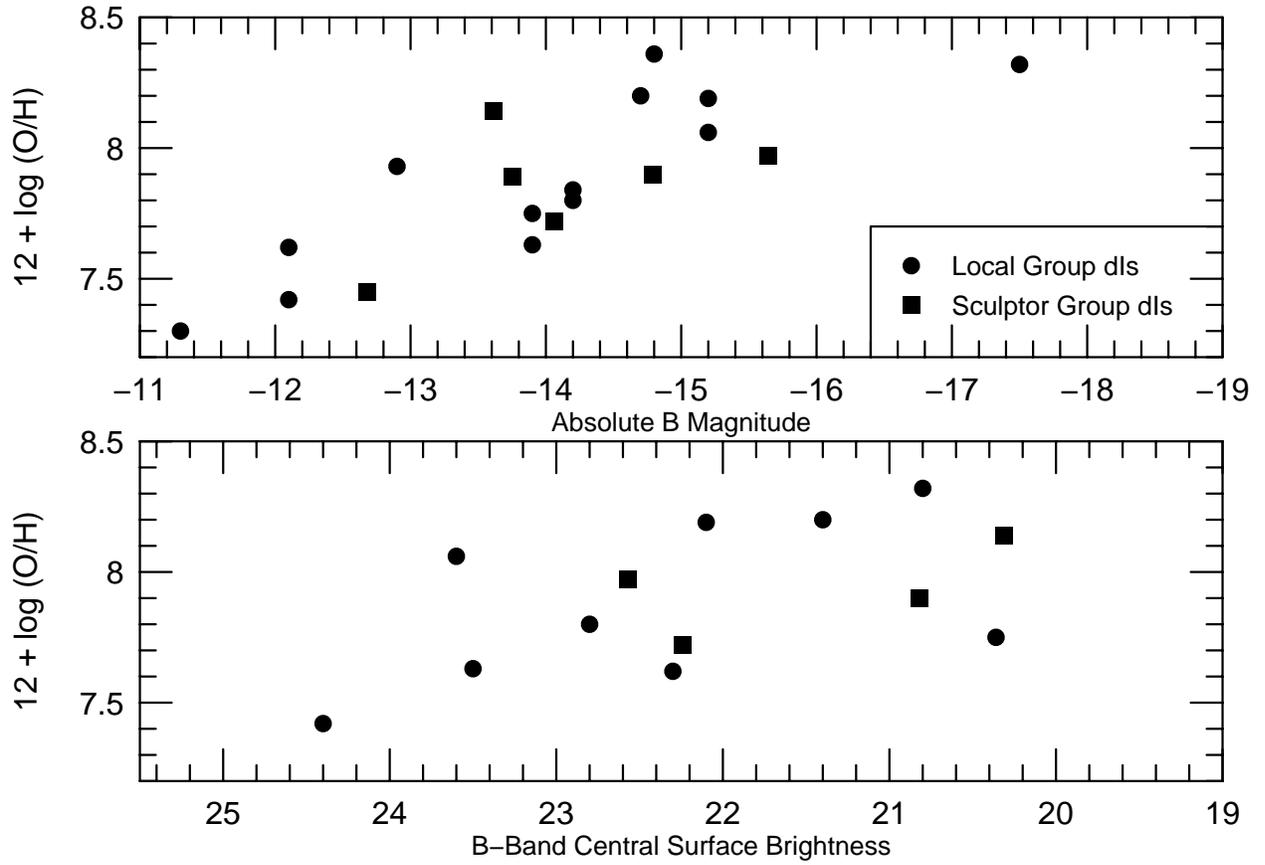}
\figcaption[k]
{Comparison of HII region oxygen abundance versus absolute
magnitude and central surface brightness for the Sculptor Group dwarf
irregular galaxies with the Local Group dwarf
irregular galaxies from the compilation of Mateo (1998).
\label{fig8}}
\end{figure}

\clearpage 

\begin {figure}
\plotone{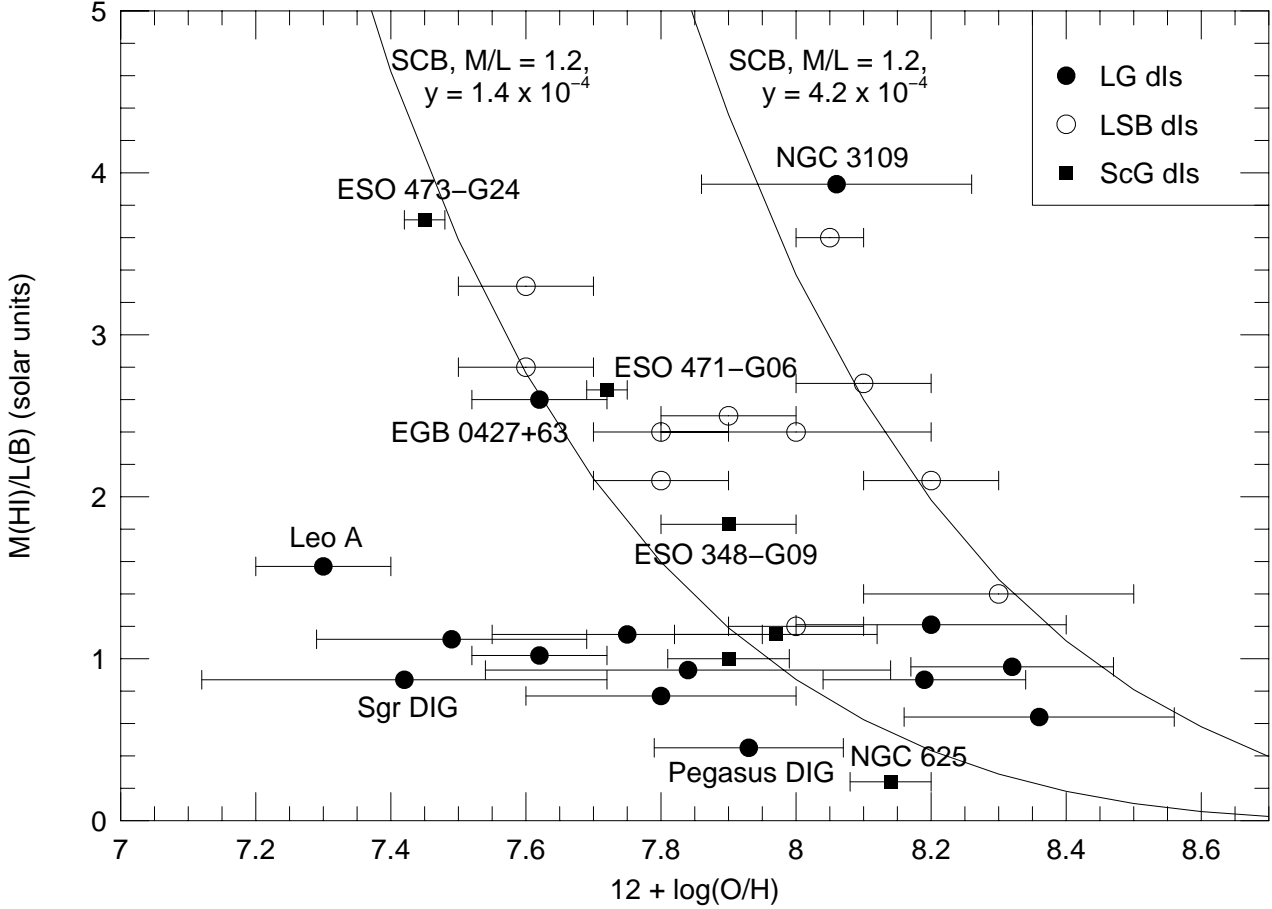}
\figcaption
{Comparison of M(HI)/L(B) versus log (O/H) for the Sculptor Group 
dwarf irregular galaxies with the Local Group dwarf
irregular galaxies from the compilation of Mateo (1998) and
LSB dwarf irregular galaxies from van Zee et al.\ (1997b).
The solid curved lines represent the evolution of simple closed box
models with the stated parameters.  The higher yield curve resulted
from a fit to DDO 154 (Kennicutt \& Skillman 2001) and is in good
agreement with theoretically predicted O yields.
\label{fig9}}
\end{figure}

\end{document}